\documentclass[lettersize,journal]{IEEEtran}
\usepackage{cite}
\usepackage{amsmath,amssymb,amsfonts}

\usepackage{algorithm} 
 
\usepackage{subfigure}
\usepackage{textcomp}
\usepackage{bm}
\usepackage{amsmath,amsfonts}
\usepackage{algorithmic}
\usepackage{array}
\usepackage[caption=false,font=normalsize,labelfont=sf,textfont=sf]{subfig}
\usepackage{textcomp}
\usepackage{stfloats}
\usepackage{url}
\usepackage{verbatim}
\usepackage{graphicx}
\usepackage{color}
\def\BibTeX{{\rm B\kern-.05em{\sc i\kern-.025em b}\kern-.08em
    T\kern-.1667em\lower.7ex\hbox{E}\kern-.125emX}}
\usepackage{balance}
\begin{document}
\title{ Energy Efficient Computation Offloading in Aerial Edge Networks With Multi-Agent Cooperation }

\author{Wenshuai Liu, Bin Li, \IEEEmembership{Member, IEEE}, Wancheng Xie, Yueyue Dai, \IEEEmembership{Member, IEEE}, \\ 
and Zesong Fei, \IEEEmembership{Senior Member, IEEE}

\thanks{Wenshuai Liu, Bin Li, and Wancheng Xie are with the School of Computer and Software,
Nanjing University of Information Science and Technology, Nanjing 210044, China
(e-mail: liuwenshuai@nuist.edu.cn; zuoyeyiwancheng@gmail.com).}

\thanks{Yueyue Dai is with the School of Cyber Science and Engineering and Research Center of 6G Mobile Communications, Huazhong University of Science and Technology, Wuhan 430074, China (e-mail: yueyuedai@ieee.org).}

\thanks{Zesong Fei is with the School of Information and Electronics, Beijing Institute of Technology, Beijing 100081, China. (e-mail: feizesong@bit.edu.cn).}
}

\maketitle

\begin{abstract}
    With the high flexibility of supporting resource-intensive and time-sensitive applications, unmanned aerial vehicle (UAV)-assisted mobile edge computing (MEC) is proposed as an innovational paradigm to support the mobile users (MUs). As a promising technology, digital twin (DT) is capable of timely mapping the physical entities to virtual models, and reflecting the MEC network state in real-time. In this paper, we first propose an MEC network with multiple movable UAVs and one DT-empowered ground base station to enhance the MEC service for MUs. Considering the limited energy resource of both MUs and UAVs, we formulate an online problem of resource scheduling to minimize the weighted energy consumption of them. To tackle the difficulty of the combinational problem, we formulate it as a Markov decision process (MDP) with multiple types of agents. Since the proposed MDP has huge state space and action space, we propose a deep reinforcement learning approach based on multi-agent proximal policy optimization (MAPPO) with Beta distribution and attention mechanism to pursue the optimal computation offloading policy. Numerical results show that our proposed scheme is able to efficiently reduce the energy consumption and outperforms the benchmarks in performance, convergence speed and utilization of resources.
\end{abstract}


\section{Introduction}
\label{sec:introduction}
\IEEEPARstart{W}{ith} the rapid advance of Internet of Things (IoT) technology and the extensive deployment of 5G networks, the envision and planning of future 6G networks is in progress.
It is expected to be empowered with the ability of providing intelligent and green communication features, and realizing the ubiquitous connection between massive objectives \cite{Nguyen2022IOTJ, Guo2021IOTJ, KB2022JSAC}. 
In order to achieve these requirements, the quality of service and application experience in 6G network need to be significantly enhanced, which involves higher requirements for the computing resources and communication resources of various devices \cite{Ji2021TII}. 
In this sense, mobile edge computing (MEC) as a popular paradigm naturally arises to alleviate the pressure on resource-constrained devices where the heterogeneous resource-intensive applications are performed. 
Specifically, the devices can offload their computational tasks to edge servers with powerful computing resources to pursue satisfactory energy consumption and latency \cite{Chen2021WN, Cao2021N}.

However, the transmission links of terrestrial MEC servers may be influenced by blockage, and the mobile users (MUs) located away from the servers are typically not well covered. Especially in some emergency situations such as disaster area and hot spot area, the fixed MEC servers are intricate to deploy.   
Recently, unmanned aerial vehicle (UAV) has emerged as a promising technology to assist traditional infrastructure-based MEC networks due to the representative features of high mobility, flexible deployment, and low cost \cite{Li2019IOTJ, Jiang2021N}. 
By cooperatively carrying MEC servers, UAVs can fly closer to the MUs to offload part of the tasks and provide reliable communication links, which is a fast and cost-effective deployment way for computation offloading with low delivery delay and high data rate requirements \cite{Liu2022TVT_Resource,Ji2022TMC}.
The dynamicity of the MEC networks is further enhanced with the participation of UAVs and MUs, and thus motivates the application of online optimizing in MEC service, especially the integration of artificial intelligence approaches that outperforms traditional offline optimization in decision time and adaptability to the environment.

For the learning-based optimizing techniques, the dynamicity in networks imposes strict requirements on accurate state sensing, real-time decision-making, and the precise execution, which presents technical challenges for implementing the UAV-assisted MEC networks \cite{Wu2021IOTJ,Mihai2022COMST}. 
In view of this, as an emerging technology in the 6G era, digital twin (DT) has come into the research vision, which is capable of maintaining the virtual models of physical entities in digital space and monitoring of the overall system status. Currently, DT has been widely used in smart city, healthcare, and intelligent manufacturing \cite{Dai2022JCN}. By collecting massive timely data from real space, DT can also realize efficient model training in dynamic MEC networks, thus providing the MEC services with more intelligent decisions.

\subsection{Related Work}

In recent years, UAV-assisted MEC has attracted growing research interests, and extensive research efforts have been conducted to enhance network performance in various scenarios. 
To exploit the performance of UAV in special situations, Do-Duy \textit{et al.} \cite{Tan2022JSAC} and Tran \textit{et al.} \cite{Tran2022TWC} studied the joint deployment and resource configuration in UAV-aided disaster emergency communications by using iterative methods. 
To pursue the real-time decision of UAV-assisted networks, Wang \textit{et al.} \cite{Wang2022TMC} integrated deep reinforcement learning (DRL) method into iterative convex optimization approaches, and
Dai \textit{et al.} \cite{Dai2022TVT} studied the minimization of the execution latency in an edge-cloud heterogeneous network comprising UAVs and base stations via DRL approach. 
For some distributed scenarios, multi-agent reinforcement learning (MARL) emerges as a mighty technique that trains efficient decentralized models, and thereby avoids the dependence of centralized DRL on the whole network state.
Wang \textit{et al.} \cite{Wang2021TCCN} considered the load balance in multi-UAV-assisted MEC under the constraints of users' energy consumption. 
Peng \textit{et al.} \cite{Peng2021JSAC} investigated the utilization of MARL to solve the dynamic network resource allocation in UAV-assisted vehicular edge networks. 
Cai \textit{et al.} \cite{Cai2021TNSE} studied the data sensing and computation offloading problem in a multi-UAV-assisted MEC, and integrated attention mechanism into MARL algorithm to accelerate the training speed of the network. 
To further investigate the cooperation of multiple agents, Ji \textit{et al.} \cite{Ji2022TMC} exploited the acquisition latency minimization in multi-UAV cellular networks via MARL. 
Zhao \textit{et al.} \cite{Zhao2022TWC} investigated the cooperation between multi-UAV and multi-ground MEC servers to minimize the total network utility and enhance the performance of service.

In view of incomparable merits of DT, the idea of combining DT and MEC has made many attempts to build a digital twin edge network (DITEN), but the current research on DITEN is still in its infancy.
To fully exploit the capability of DT network for capturing the real-time state of the network, 
Sun \textit{et al.} \cite{Sun2020TVT} investigated the latency minimization problem of DITEN via DRL method and considered the estimated deviation of DT.
Sun \textit{et al.} \cite{Sun2021TII} developed an approach to automatically adjust the aggregation frequency of federated learning in DITEN network.
Lu \textit{et al.} \cite{Lu2021TII} proposed a blockchain-based federated learning framework that runs in DT for collaborative computing, improving the reliability and security of the system and enhancing data privacy.
Liu \textit{et al.} \cite{Liu2022IOTJ} introduced edge collaboration into the DITEN network and realized intelligent offloading of tasks.
As a step forward, Do-Duy \textit{et al.} \cite{Tan2022IWCL} iteratively optimized the joint resources of mobile devices and multiple edge servers to minimize the total latency in industrial IoT networks, and Zhang \textit{et al.} \cite{Zhang2022TII} combined DT technology and decomposed MARL into the design of vehicle edge computing networks. Recently, Li \textit{et al.} \cite{Li2022DT} focused on the DT-aided computation offloading in UAV edge networks where the double deep Q-network algorithm is explored to reduce the system energy consumption.

\subsection{Contributions and Organization}
The above-mentioned excellent works have investigated the UAV/DT networks from various aspects. 
However, the scenarios that involve the joint cooperation among UAVs,  base station (BS), and MUs have not been fully exploited. Especially when MUs are deployed in remote or disaster areas, they are not able to directly communicate and synchronize status with BS and DT. Therefore, different from these prior studies, we in this paper propose a real-time distributed optimization framework for the  multi-UAV-assisted MEC networks based on MARL, where the policy model can be deployed on MUs and UAVs to make resource allocation decisions, and a DT layer is deployed at the BS to pursue timely monitoring and centralized training, thereby saving the energy of model training on resource-limited MUs and UAVs. Particularly, the main contributions of this paper are presented as follows. 
\begin{enumerate}
    \item We proposed a multi-UAV-assisted MEC network with air-ground cooperation to provide the computing services for the resource-limited MUs, where the interplay between physical environment and DT layer is taken into account. The MUs can partially offload their computational task to UAVs or relay to BS for further computation processing via UAVs. Considering the energy limits of both MUs and UAVs, a weighted energy consumption minimization problem is formulated by jointly designing the association, offloading proportion, trajectory control, and resource allocation on computation and communication. 
    \item To tackle the high complexity of the formulated real-time problem, we formulate it as a Markov decision process (MDP), and address it by an MARL framework with heterogeneous agents.  the high-dimensional hybrid action spaces, the multi-agent proximal policy optimization (MAPPO) algorithm is utilized to train both MUs and UAVs to cooperatively make offloading decisions in dynamic MEC networks. 
    \item To enhance the performance of training, we adopt Beta distribution and attention mechanism in actor and critic network, respectively. Simulation results show the efficient training convergence and effectiveness of our proposed MAPPO with attention mechanism and Beta distribution (AB-MAPPO) algorithm in optimizing energy consumption. The proposed algorithm outperforms other benchmarks, especially in utilization of computational resource, and flexibility to different deploying scenarios. 
\end{enumerate}

The rest of this paper is organized as follows. We first present the system model and formulate the joint problem in Section \ref{s:sys}. The design of AB-MAPPO algorithm is given in Section \ref{s:proposed}. Section \ref{s:simulation} presents the simulation results. Finally, we conclude this paper in Section \ref{s:conclusion}.

\begin{figure*}[t]
    \centering
    \includegraphics[width=\textwidth]{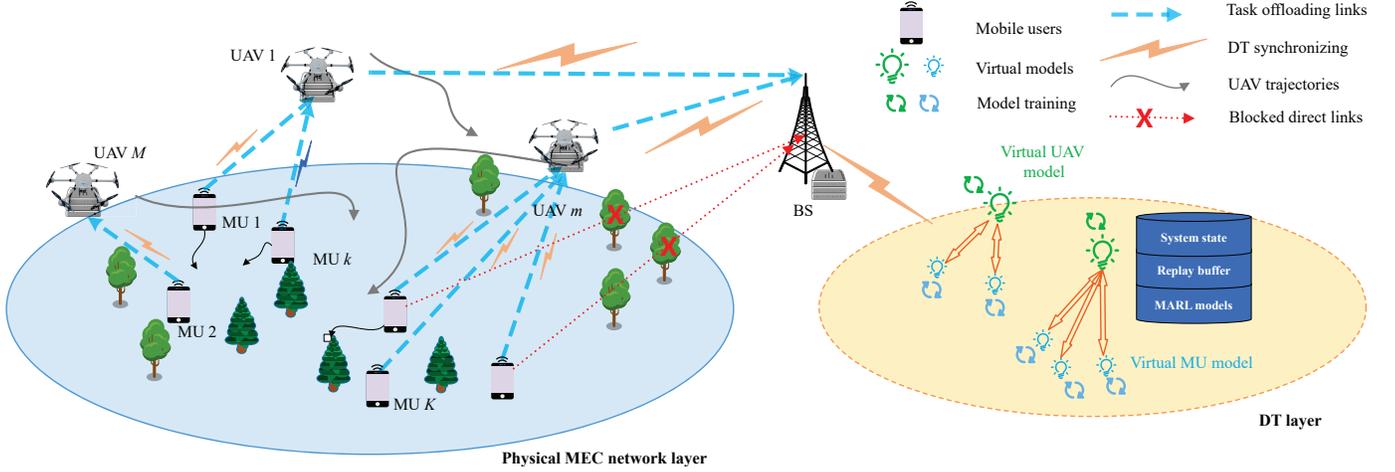}
    \caption{System model.}
    \label{fig:sys}
\end{figure*}

\section{System Model and Problem Formulation}\label{s:sys}
{In this section, we first introduce a multi-UAV-assisted MEC system model, including network model, mobility model of MUs, DT model, communication model, and computation model. Then, we formulate the optimization problem to minimize the weighted energy consumption of MUs and UAVs while ensuring the delay requirement of the tasks.}
 
\subsection{Network Model}
As shown in Fig. \ref{fig:sys}, we consider a multi-UAV-assisted air-ground MEC network containing a BS, $M$ UAVs, and $K$ MUs. For simplicity, we denote the set of MUs by $\mathcal{K}\triangleq\{1,2,\dots,K\}$, the set of UAVs by $\mathcal{M}\triangleq\{1,2,\dots,M\}$, and the index of BS by $M+1$. 
The MUs typically have limited battery life and computing capability, which are not capable of completing their resource-intensive tasks efficiently. 
We assume that UAVs and BS are equipped with MEC servers to provide computing service for MUs. 
The set of servers is denoted by $\mathcal{M}^\star \triangleq \mathcal{M}\cup\{M+1\}$. Without loss of generality, we introduce a time period $T$, which is divided into $N$ time slots with time of $\delta_t$. 
The set of time slots is denoted as $\mathcal{N}\triangleq\{1,2,\dots,N\}$. 
Using three-dimensional Cartesian coordinate system, the locations of MU $k$, UAV $m$ and BS at time slot $n$ are denoted as ${\bf u}_k[n]=[x_k^{\rm MU}[n],y_k^{\rm MU}[n],0]^{\text{T}}$, ${\bf q}_m[n]=[x_m^{\rm UAV}[n],y_m^{\rm UAV}[n],H]^{\text{T}}$, and ${\bf u}_{\text{BS}}=[x_{\text{BS}},y_{\text{BS}},H_\text{B}]^{\text{T}}$, respectively. 
Note that the UAVs are flying at a fixed altitude $H$. To establish an efficient mapping of the physical MEC network, a DT layer is deployed at BS and is equipped with essential functions of data storage, status synchronization and model training. 
The main notations in this paper are summarized in Table \ref{tab:notations}.

\subsection{Mobility Model of MUs}
At the beginning of time slot $n=1$, all MUs are randomly located, and the MUs are moving according to Gaussian-Markov random model \cite{Liu2020TVT}. Considering that each time slot is of short period, the location of MUs are assumed to be static during one time slot. Specifically, at each time slot $n$, the velocity of $v_k[n]$ and direction $\theta_k[n]$ of MU $k$ are respectively given by
\begin{subequations}
    \begin{align}
        &v_k[n]=\mu_1v_k[n-1]+(1-\mu_1)\bar{s} + \sqrt{1-\mu_1^2}\Phi_k, \\
        &\theta_k[n]=\mu_2\theta_k[n-1]+(1-\mu_2)\bar{\theta} + \sqrt{1-\mu_2^2}\Psi_k,
    \end{align}
\end{subequations}
where $0\leq \mu_1,\mu_2 \leq 1$ are the parameters representing the effect of previous state, $\bar{s}$ and $\bar{\theta}$ are the average velocity and direction of all MUs, respectively. Also, $\Psi_k$ and $\Phi_k$ are generated by two independent Gaussian distributions with different mean-variance pairs $(\bar{\xi}_{v_k}, \varsigma^2_{v_k})$ and $(\bar{\xi}_{\theta_k}, 
\varsigma^2_{\theta_k})$ for MU $k$. Therefore, the coordinate of MUs can be updated as
\begin{subequations}
    \begin{align}
        &x_k^{\rm MU}[n]=x_k^{\rm MU}[n-1]+v_k[n-1]\cos(\theta_k[n-1])\delta_t,\\
        &y_k^{\rm MU}[n]=y_k^{\rm MU}[n-1]+v_k[n-1]\sin(\theta_k[n-1])\delta_t.
    \end{align}
\end{subequations}

{
\subsection{DT Model of the UAV-assisted MEC Network}
In the considered multi-UAV network with MEC service, the DT layer deployed at BS is responsible for state monitoring and virtual twin mapping.
To maintain the virtual twins, the physical devices need to upload the DT information of themselves to the DT layer. In this paper, two types of entities are represented, i.e., MUs and the UAVs. To model the virtual twins of them, we focus on their key features which efficiently represent their real-time state corresponding to the optimization scheme.
}

{
The virtual twin of each MU $k$ needs to record its location and task information at each time slot $n$, which is given by
\begin{align}
    {\rm DT}_k[n]=\{{\bf u}_k[n],\Omega_k[n],\tilde{f}_k^{\text{loc}}[n]\},
\end{align}
where $\Omega_k[n]$ is the task information of users and $\tilde{f}_k^{\text{loc}}[n]$ is the estimated computational frequency for MU $k$, which will be illustrated in following subsection. 
}

{
The DT of each UAV should reflect its service status, involving the resource allocation and movement, and is denoted as
\begin{align}
    {\rm DT}_m^U[n]=\{{\bf q}_m[n],\alpha_{k,m}[n],\tilde{f}_{k,m}[n],\forall k\in\mathcal{K}\},
\end{align}
where $\alpha_{k,m}[n]$ and $\tilde{f}_{k,m}^{\text{edge}}[n]$ are defined the association factor of the network and the estimated computational resource allocated to MU $k$ by UAV $m$.
}
\begin{table*}[t]
    \caption{{Main Notations}}\centering
     \renewcommand{\arraystretch}{1.5}
    \begin{tabular}{|l|p{2.3in}|p{0.5in}|p{2.3in}|}
    \hline
    \textbf{Notation}                             & \textbf{Definition}                                                                  & \textbf{Notation}                      & \textbf{Definition}                                                    \\ \hline
    $K$                                           & The number of MUs                                                                    & $M$                                    & The number of UAVs                                                     \\ \hline
    $N$                                           & The number of time slots per period                                                  & $B$                                    & The total bandwidth of the network                                     \\ \hline
    $\mathcal{M}$                                 & The set of UAVs                                                                      & $\mathcal{M}^\star$                    & The set of MEC servers                                                 \\ \hline
    $\mathcal{K}$                                 & Offloading proportion of MU $k$ at time slot $n$                                     & $\mathcal{N}$                          & The set of time slot                                                   \\ \hline
    ${\bf q}_m[n]$                    & The location of MEC server $m$ at time slot $n$                                      & 
    ${\bf u}_k[n]$              & The location of MU $k$ at time slot $n$                                \\ \hline
    ${\bf u}_{\text{BS}}$                       & The location of BS                                                & $L_k[n]$                               & The size of input data of MU $k$                                       \\ \hline
    $\rho_k[n]$                                   & The offloading proportion of MU $k$                                                  & $T_{k,m}^{\text{ecmp}}[n]$             & The actual edge computing time of MU $k$                               \\ \hline
    $C_k[n]$                                      & Number of CPU cycles required for processing one bit data of MU $k$ at time slot $n$ & $\alpha_{k,m}[n]$                      & The association variable between MU $k$ and UAV $m$ at time slot $n$   \\ \hline
    $f_{k,m}^{\text{edge}}[n]$                    & The actual allocated CPU frequency of UAV $m$ to MU $k$ at time slot $n$                    & $f_k^{\text{loc}}[n]$                  & The actual CPU frequency of MU $k$ at time slot $n$                           \\ \hline
    $T_k^{\text{loc}}[n]$, $T_k^{\text{edge}}[n]$ & The actual local computing time and edge computing time of MU $k$                    & $B_{k,m}[n]$                           & The bandwidth allocated to link of MU $k$ and UAV $m$ at time slot $n$ \\ \hline
    $\bm{v}_m[n]$, ${\bf a}_m[n]$                 & The acceleration and velocity of UAV $m$ at time slot $n$                            & $s(t)$,\;$o_i(t)$, $a_i(t)$,\;$r_i(t)$ & The global state, observation and reward of agent $i$ at time step $t$ \\ \hline
    \end{tabular}
\label{tab:notations}
\end{table*}

\subsection{Computation Model}\label{subsec:comp}
Each MU generates a task $\Omega_k[n]=(L_k[n], C_k[n])$ with a latency requirement of $\delta_t$ at the beginning of each time slot $n$, where $L_k[n]$ and $C_k[n]$ denote the size of input data and average number of central process unit (CPU) cycles for processing one bit data of MU $k$'s task, respectively. 
The tasks can be divided into two parts at random and executed concurrently, with $(1-\rho_k[n])L_k[n]$ bits of input data for local computing, and $\rho_k[n]L_k[n]$ bits offloaded to MEC servers associated with BS and UAVs for edge computing. 
Denoting the offloading association of MU $k$ at time slot $n$ as $\alpha_{k,m}[n]$, we have
\begin{equation}
    \alpha_{k,m}[n]\in\{0,1\},\forall k \in\mathcal{K}, m \in\mathcal{M}^\star, n \in\mathcal{N}.\label{c:alpha_zeroone}
\end{equation}
The details are illustrated as follows.
\subsubsection{Local computing} With $\alpha_{k,0}[n]=1$, MU $k$ computes total task by itself locally, indicating that $\rho_k[n]=0$. 

According to prior MEC research \cite{Dai2022TVT}, {by adopting dynamic voltage frequency scaling (DVFS) technique}, the estimated frequency of MU $k$ at time slot $n$ in DT can be denoted as 
\begin{equation}
    \tilde{f}_k^{\text{loc}}[n]=\min\{{ Y_k^{\text{loc}}[n]/ \delta_t}, f^{\text{loc}}_{\max}\},
\end{equation}
where $f^{\text{loc}}_{\max}$ denotes the maximum CPU frequency of MUs, $Y_k^{\text{loc}}[n]=(1-\rho_k[n])L_k[n] C_k[n]$ denotes the total computing cycles of MU $k$'s task.
It means that for energy efficient objective, the task is processed with the minimum CPU frequency required to complete it according to DVFS, thus we have $\tilde{f}_k^{\text{loc}}[n]= Y_k^{\text{loc}}[n] / \delta_t$, and $\tilde{f}_k^{\text{loc}}[n]$ is limited by the maximum CPU frequency $f^{\text{loc}}_{\max}$. 

It is noteworthy that the DT layer can't fully reflect the state of MUs and UAVs due to some issues such as the hysteresis of status synchronization and the transmission error. Thus, a deviation is introduced to model the estimated error of DT and can be utilized to verify the robustness of the system. The deviation may occur from computational frequency \cite{Sun2020TVT, Liu2022IOTJ, Tan2022IWCL}, and location data.   Denoting $\tilde{f}_k^{\text{loc}}[n]$ as the estimated CPU frequency of MU $k$, the estimated local computing time of MU $k$ at time slot $n$ can be expressed as $\tilde{T}_k^{\text{loc}}[n]=Y_k^{\text{loc}}[n]/\tilde{f}_k^{\text{loc}}[n]$. By denoting $\hat{f}_{k}^{\text{loc}}[n]$ as the estimated deviation of actual frequency $f_{k}^{\text{loc}}[n]=\tilde{f}_{k}^{\text{loc}}[n]+\hat{f}_{k}^{\text{loc}}[n]$, the computing latency gap of MU $k$ between DT and actual value can be expressed as
\begin{equation} 
        \Delta T_k^{\text{loc}}[n]=\frac{-Y_k^{\text{loc}}[n]{\hat{f}_k^{\text{loc}}[n]}}{\tilde{f}_k^{\text{loc}}[n](\tilde{f}_k^{\text{loc}}[n]+{\hat{f}_k^{\text{loc}}[n]})}.
\end{equation}

As a result, the actual local computing time of MU $k$ can be calculated by 
\begin{equation}
    T_k^{\text{loc}}[n]=\tilde{T}_k^{\text{loc}}[n]+\Delta T_k^{\text{loc}}[n].
\end{equation}

\subsubsection{Edge computing} MUs can request UAVs for offloading their tasks to MEC servers at UAVs or BSs. The procedures are respectively illustrated as follows. 
\begin{itemize}
    \item \textbf{UAV computing}: A part of MU $k$'s task is transmitted to UAV $m$, and executed by the MEC server on UAV. In this case, we have $\alpha_{k,m}[n]=1$ and $\alpha_{k,M+1}[n]=0$.
    \item \textbf{BS computing}: Considering the complex ground environment such as obstacle blocking, the MU-BS links are too poor to support transmission. Therefore, a certain part of task is first transmitted to UAV $m$, and further relayed to BS for executing. To this end, we have $\alpha_{k,m}[n]=\alpha_{k,M+1}[n]=1$.
\end{itemize}

We adopt probabilistic line-of-sight (LoS) channel to represent UAV-MU links. The probability of geometrical LoS between the UAV and MUs depends on the environment status and the elevation angle. Denoting the elevation angle as $\vartheta_k[n]$, and the LoS probability of MU $k$ and UAV $m$ at time slot $n$ as $\mathbb{P}(\text{LoS},\vartheta_k[n])$, which is approximated to following form
\begin{equation}
    \mathbb{P}(\text{LoS},\vartheta_k[n])=\frac{1}{1+a\text{exp}\left(-b(\vartheta_k[n]-a)\right)},
\end{equation}
where $a$ and $b$ are the parameters related to environment, and $\vartheta_k[n]$ is given by
\begin{equation}
    \vartheta_k[n]=\frac{180}{\pi}\arctan\left(\frac{H}{\Vert{\bf u}_k[n]-{\bf q}_m[n]\Vert}\right).
\end{equation}

Additionally, the non-line-of-sight (NLoS) channel probability can be expressed as $\mathbb{P}(\text{NLoS},\vartheta_k[n])=1-\mathbb{P}(\text{LoS},\vartheta_k[n])$.
Therefore, the expected channel power gain of from MU $k$ to UAV $m$ is given by
\begin{align}
        \nonumber h_{k,m}[n]&=\frac{\beta_0\mathbb{P}(\text{LoS},\vartheta_k[n])}{\Vert {\bf u}_k[n]-{\bf q}_m[n]\Vert^{\tilde{\iota}}}+\frac{\nu \beta_0\mathbb{P}(\text{NLoS},\vartheta_k[n])}{\Vert {\bf u}_k[n]-{\bf q}_m[n]\Vert^{\tilde{\iota}}}\\
        &=\frac{\nu \beta_0\mathbb{P}(\text{NLoS},\vartheta_k[n])}{\Vert {\bf u}_k[n]-{\bf q}_m[n]\Vert^{\tilde{\iota}}},
\end{align}
where $\tilde{\iota}$ is the path loss exponent, $\nu$ is the NLoS attenuation, $\beta_0$ is the channel power gain at the reference distance of 1 m. Similar to \cite{Yang2022TWC}, we assume that the change of LoS probability $\mathbb{P}(\text{LoS},\vartheta_k[n])$ in each time slot is negligible.
Furthermore, channel from UAV $m$ to BS can be modeled by quasi-static block fading LoS link \cite{Xu2021TWC}, i.e.,
\begin{equation}
h_m^{\text{rel}}[n]=\frac{\beta_0}{\Vert {\bf q}_m[n]-{\bf u}_{\text{BS}}\Vert^2}.
\end{equation} 

We consider the orthogonal frequency division multiple access scheme for data transmission. In each time slot, MU $k$ first requests for offloading, then UAV $m$ allocates an orthogonal frequency bandwidth of $B_{k,m}[n]$ for transmission. Therefore, the transmit rates between MU $k$ and UAVs, and UAV-BS are given by
\begin{equation}
    R_k[n]= \sum\limits_{m=1}^M \alpha_{k,m}[n] B_{k,m}[n]\log_2 \left(1+\frac{p_k h_{k,m}[n]}{B_{k,m}[n]N_0}\right),\label{eq:R}
\end{equation}
\begin{equation}
\begin{split}
    R_{m}^{\text{rel}}[n] = B_u\log_2\left(1+\frac{p_m h_m^{\text{rel}}[n]}{B_u N_0}\right),
    \label{eq:R_BS}
\end{split}
\end{equation}
where $p_k$ and $p_m$ denote the uplink transmit power of MU $k$ and UAV $m$, respectively, $B_u$ is the bandwidth allocated to UAVs for relaying, and $N_0$ is the noise power density.

Since each MU can associate with at most one UAV or BS in each time slot $n$, and the bandwidth should be only allocated to links between associated devices, we have
\begin{gather}
    \sum\limits_{m=0}^{M}\alpha_{k,m}[n]= 1, \forall k\in \mathcal{K}, n \in \mathcal{N},\label{c:alpha_1}\\
    \sum\limits_{m=1}^{M}\sum\limits_{k=1}^{K} B_{k,m}[n] \leq B, \forall n \in \mathcal{N},\label{c:band_alloc}\\
    B_{k,m}[n] \ge 0, \forall k\in \mathcal{K},\forall m \in \mathcal{M},  n \in \mathcal{N},\label{c:Bge0}\\
    \lceil B_{k,m}[n]/B \rceil =\alpha_{k,m}[n],\forall k\in \mathcal{K}, m \in \mathcal{M},  n \in \mathcal{N}\label{c:band_alloc_on},
\end{gather}
where $B$ denotes the available channel bandwidth. Then the size of total task transmitted to UAV $m$ by MUs and the CPU cycles of MU $k$'s task executed by UAV $m$ can be respectively calculated as
\begin{gather}
    L_m^{\text{trans}}[n]=\sum\limits_{k=1}^{K}\alpha_{k,m}[n]\rho_k[n]L_k[n],\\
    Y_{k,m}^{\text{edge}}[n]=\alpha_{k,m}[n](1-\alpha_{k,M+1}[n])\rho_k[n]L_k[n]C_k[n].
\end{gather}

Following prior research \cite{Zhao2021JSAC,Dai2022TVT}, we assume that the computation result with small size can be downloaded to MU with a negligible transmission latency. Additionally, BS is equipped with energy supply and powerful MEC server, hence the computing time and energy at BS are not considered.
Thus, the total estimated edge computing latency of MU $k$ can be expressed as 
\begin{align}
    \nonumber T_k^{\text{edge}}[n]=&\sum\limits_{m=1}^M\alpha_{k,m}[n]\left(T_k^{\text{off}}[n]+\alpha_{k,M+1}[n]T_{k,m}^{\text{rel}}[n]\right.\\ 
    &\left.+(1-\alpha_{k,M+1}[n])T_{k,m}^{\text{ecmp}}[n]\right),
\end{align}
where 
$T_k^{\text{off}}[n]=L_k[n]\rho_k[n]/R_{k,m}[n]$ denotes the transmission delay of MU $k$, 
$T_m^{\text{rel}}[n]=L_k[n]\rho_k[n]/R_{k,m}^{\text{rel}}[n]$ denotes the time of UAV $m$ for relaying the task to BS,
 and $\tilde{f}_{k,m}^{\text{edge}}[n]$ is the estimated CPU frequency of server $m$ allocated to MU $k$ at time slot $n$. Accordingly, the computing latency gap of UAV $m$ between DT and actual value can be expressed as
\begin{equation} 
        \Delta T_{k,m}^{\text{ecmp}}[n]=\frac{-Y_{k,m}^{\text{edge}}[n]{\hat{f}_{k,m}^{\text{edge}}[n]}}{\tilde{f}_{k,m}^{\text{edge}}[n](\tilde{f}_{k,m}^{\text{edge}}[n]+{\hat{f}_{k,m}^{\text{edge}}[n]})},
\end{equation}
where $\hat{f}_{k,m}^{\text{edge}}[n]$ is the estimated deviation of $f_{k,m}^{\text{edge}}[n]$. Denoting $\tilde{T}_{k,m}^{\text{ecmp}}[n]=Y_{k,m}^{\text{edge}}[n]/\tilde{f}_{k,m}^{\text{edge}}[n]$
as the estimated time of UAV $m$ for computing the task of MU $k$, the actual edge computing time of MU $k$ at each time slot $n$ can be calculated by 
\begin{equation}
    T_{k,m}^{\text{ecmp}}[n]=\tilde{T}_{k,m}^{\text{ecmp}}[n] + \Delta T_{k,m}^{\text{ecmp}}[n],
\end{equation}
and the actual CPU frequency allocated to MU $k$ by UAV $m$ is given by 
    $f_{k,m}^{\text{edge}}[n]=\tilde{f}_{k,m}^{\text{edge}}[n]+\hat{f}_{k,m}^{\text{edge}}[n]$.
Since it is inappropriate to allocate the CPU frequency to MUs which are not choosing to offload to the UAVs, i.e., computing at local or BS, we have
\begin{align}
\nonumber
    \lceil \tilde{f}_{k,m}^{\text{edge}}[n]/f_{\max}^{\text{edge}} \rceil =&\alpha_{k,m}[n](1-\alpha_{k,M+1}[n]), \\
    &\forall k \in \mathcal{K}, m \in \mathcal{M}, n \in \mathcal{N}\label{c:freq_alloc_on}.
\end{align}

{For the location data, a random noise is imposed on the observation of devices, and thus has an influence on the decisions of algorithm, which will illustrated in Section \ref{s:proposed}.}

\subsection{Energy Consumption Model}
As the entities (MUs and UAVs) can upload their essential status information to DT layer via UAVs with specific bandwidth, the energy consumption can be evaluated by the DT layer. The energy models of MUs and UAVs are illustrated as follows.

\subsubsection{MU energy consumption} The energy consumption of MU $k$ comprises the local computing energy and transmit energy, which are denoted as $E_k^{\text{loc}}[n]$ and $E_k^{\text{off}}[n]$, respectively. Therefore, we have
\begin{equation} 
    E_k^{\text{loc}}[n]=\kappa f_k^{\text{loc}}[n]^2 Y_k^{\text{loc}}[n],
\end{equation}
\begin{equation} 
    E_k^{\text{off}}[n]=p_k[n]T_k^{\text{off}}[n],
\end{equation}
where $\kappa$ is the effective capacitance coefficient determined by chip architecture.
\subsubsection{UAV energy consumption} The energy consumption of UAVs is imposed by flying and computation. In each time slot $n$, UAV $m$ is assumed to fly under the limits of the maximum speed $v_{\max}$ and the maximum acceleration $a_{\max}$, which indicates that 
{
\begin{equation}
    \Vert{\bf a}_m[n]\Vert=\frac{\Vert{\bf v}_m[n+1]-{\bf v}_m[n]\Vert}{\delta_t}\leq a_{\max},
     \forall m\in \mathcal{M}, n \in \mathcal{N},\label{c:a}
\end{equation}
}
\begin{equation}
    {\bf q}_m[n+1]={\bf q}_m[n]+{\bf v}_m[n]\delta_t+\frac{1}{2}{\bf a}_m[n]\delta_t^2,\forall m\in \mathcal{M}, n \in \mathcal{N},\label{c:u}
\end{equation}
\begin{equation}
    \Vert{\bf v}_m[n]\Vert = \frac{\Vert{\bf q}_m[n+1]-{\bf q}_m[n]\Vert}{\delta_t} \leq v_{\max},\forall m\in \mathcal{M}, n \in \mathcal{N}\label{c:v}.
\end{equation}
Subsequently, we adopt the UAV's propulsion energy model introduced by \cite{Zeng2019TWC}, the flying energy of UAV $m$ is then expressed as follows
\begin{align}
         E_m^{\text{fly}}[n]=&\Bigg[\frac{1}{2}d_0\rho s A \Vert{\bf v}_m[n]\Vert^3+P_0\left(1+\frac{3\Vert {\bf v}_m[n]\Vert^3}{U_{\text{tip}}^2}\right) \nonumber\\
         &+P_i\left(\sqrt{1+\frac{\Vert{\bf v}_m[n]\Vert^4}{4v_0^4}}-\frac{\Vert{\bf v}_m[n]\Vert^2}{2v_0^2}\right)\Bigg]\delta_t,
         \label{energy_fly}
\end{align}
where $P_0$ and $P_i$ are the blade profile power and induced power in hovering status, respectively, $U_{{\text{tip}}}$ is the tip speed of the rotor blade, $v_0$ is the mean rotor velocity, $d_0$ is fuselage drag ratio, $s$ denotes the rotor solidity, $\rho$ denotes the air density, and $A$ denotes the rotor disc area.

{Note that the communication energy of UAV is relatively small compared to the energy consumption imposed by the intensive computation and flying, which can be neglected.} The computation energy of UAV $m$ is calculated as
\begin{equation}
    E_m^{\text{edge}}=\sum\limits_{k=1}^{K}\kappa \alpha_{k,m}[n] f_{k,m}^{\text{edge}}[n]^2  Y_{k,m}^{\text{edge}}[n].
\end{equation}

As a consequence, the overall energy consumption of MUs and UAVs are calculated by
\begin{equation}
    E_k^{\text{MU}}[n] = E_k^{\text{loc}}[n]+E_k^{\text{off}}[n],
\end{equation}
\begin{equation}
    E_m^{\text{UAV}}[n] = E_m^{\text{fly}}[n]+E_m^{\text{edge}}[n].
\end{equation}
\subsection{Problem Formulation}
Considering the deficient energy resource of MUs and limited energy budget of UAVs, we pursue a multi-UAV-assisted MEC network with air-ground cooperation design, aiming at minimizing the weighted energy consumption of MUs and UAVs with latency sensitive tasks, by jointly optimizing the MU association $\textbf{A} \triangleq \{\alpha_{k,m}[n],\forall n \in \mathcal{N},  k \in \mathcal{K},  m \in \mathcal{M}^\star\cup\{0\} \}$, the offloading proportion $\bm{\varrho}\triangleq\{\rho_k[n],\forall n \in \mathcal{N},  k \in \mathcal{K}\}$, the CPU frequency allocation $\textbf{F}\triangleq\{\tilde{f}_{k,m}^{\text{edge}}[n], \forall m \in \mathcal{M},k \in \mathcal{K}, n \in \mathcal{N}\}$, the bandwidth allocation $\textbf{B}\triangleq \{B_{k,m}[n],\forall m \in \mathcal{M},k \in \mathcal{K}, n \in \mathcal{N}\}$, and the UAV velocity $\textbf{V}\triangleq \{{\bf v}_m[n],\forall m \in \mathcal{M},n \in \mathcal{N}\}$. Therefore, the weighted energy minimization problem is formulated as 
{
\begin{subequations}
    \begin{align}
        \mathcal{P}1&\min\limits_{\textbf{A},\textbf{B},\textbf{F},\bm{\varrho},\textbf{V}} \quad \varpi\sum\limits_{n=1}^N\sum\limits_{m=1}^M E_m^{\text{UAV}}[n]+\sum\limits_{n=1}^N\sum\limits_{k=1}^KE_k^{\text{MU}}[n]\label{obj}\\
        \text{s.t}\quad&\eqref{c:alpha_zeroone},\eqref{c:alpha_1},\eqref{c:band_alloc},\eqref{c:Bge0},\eqref{c:band_alloc_on},\eqref{c:freq_alloc_on},\eqref{c:a},\eqref{c:u},\eqref{c:v},\\
        &0\leq\rho_k[n]\leq 1,\forall k \in \mathcal{K}, n \in \mathcal{N},\label{c:rho_zeroone}\\
        &\lceil\rho_k[n]\rceil=1-\alpha_{k,0}[n],\forall k \in \mathcal{K}, n \in \mathcal{N},\label{c:rho_alpha}\\
        &\sum\limits_{k=1}^K f_{k,m}^{\text{edge}}[n] \leq f_{\max}^{\text{edge}},\forall m \in \mathcal{M}, n \in \mathcal{N},\label{c:f_max}\\
        & {T}_k^{\text{loc}}[n]\leq\delta_t, \forall k \in \mathcal{K}, \forall m \in \mathcal{M}, n \in \mathcal{N},\label{c:latency_loc}\\
        &{T}_k^{\text{edge}}[n] \leq \delta_t,\forall k \in \mathcal{K}, \forall m \in \mathcal{M}, n \in \mathcal{N}, \label{c:latency_edge}\\
        &\Vert {\bf q}_i [n] - {\bf q}_j [n] \Vert^2 \ge d_{\rm{min}}^2,\forall i,j \in \mathcal{M},i \ne j,n \in \mathcal{N},\label{c:se}
    \end{align}
\end{subequations}
}
where $\varpi$ is the weight factor, $d_{\rm{min}}$ denotes the minimum safe distance between the UAVs. Constraints \eqref{c:alpha_zeroone} and \eqref{c:alpha_1} ensure the feasibility of the association status, constraints \eqref{c:band_alloc}-\eqref{c:band_alloc_on} guarantee that the bandwidth is only allocated to valid links, constraints \eqref{c:a}-\eqref{c:v} denote the velocity and acceleration limits of UAVs, constraints \eqref{c:rho_zeroone} and \eqref{c:rho_alpha} indicate the range and the validness of offloading proportion, constraints \eqref{c:freq_alloc_on} and \eqref{c:f_max} denotes feasibility of computational resource allocation, constraints \eqref{c:latency_loc} and \eqref{c:latency_edge} denote the computational latency constraints, and constraint \eqref{c:se} limits the minimum safe distance between UAVs.

It can be readily derived that $\mathcal{P}1$ is a mixed integer non-convex and combinational problem with huge amount of highly decoupled variables. Furthermore, the proposed problem needs to be timely solved in each time slot $n$ due to the randomly generated tasks and the movement of devices. Nonetheless, it is very intricate to solve the problem by typical iterative techniques such as alternative optimization and genetic algorithm. In the following, we propose an efficient algorithm for this complicated problem by harvesting the MARL approach.

\section{Proposed DRL Approach: AB-MAPPO}\label{s:proposed}

In this section, the essential elements of MDP in multi-agent DRL are first illustrated. Then, we endeavor to develop an MAPPO approach with attention mechanism and Beta distribution to address problem $\mathcal{P}1$.
\subsection{MDP Elements Formulation}
In MARL setup, problem $\mathcal{P}1$ can be modeled as an MDP with the set of agents $\mathcal{I}\triangleq\{1,\dots,K+M\}$, state space  $\mathcal{S}$, action space $\mathcal{A}=\mathcal{A}_1\times\mathcal{A}_2\times\dots\times\mathcal{A}_{K+M}$.
The agents can enhance their policy by interacting with the environment in discrete steps. 
In each step $t$, each agent $i$ obtains the current observation $o_i(t)$ from the global environment state $s(t)\triangleq \{o_i(t),\forall i \in \mathcal{I}\}$, takes action $a_i(t)\in\mathcal{A}_i$, then obtains a reward $r_i(t)$, and the environment transfers to a new state $s(t+1)$. 

The MDP involve two types of agents, i.e., MUs and UAVs. At the beginning of each step $t$, the MUs generate actions, and then request for offloading according to the actions. Afterwards, the UAVs obtain the actions and serve the MUs. The status information is periodically synchronized to DT if is required, and rewards are evaluated centrally by DT. The formulation of MDP elements are illustrated as follows.

\subsubsection{MDP elements of MUs} The MDP elements of MUs involve the observation $o_k(t)$, action $a_k(t)$ and reward $r_k(t)$, which are presented as follows.
\begin{itemize}
    \item \textbf{Observation:}
    {
     From the perspective of privacy, MUs can only obtain the positions of themselves ${\bf u}_k[n]$ by Global Positioning System, and the positions of MEC servers ${\bf q}_m[n]$ from the broadcast of the DT layer. The position data accordingly has a random deviation, which is also imposed to the observations of agents.} Furthermore, the task information $\Omega_k[n]$ can be observed. As the task load on UAVs is highly relevant to MUs' offloading decisions, the previous computation load of UAVs are also involved. Therefore, the observation of MU $k$ at step $t$ can be expressed as 
    \begin{equation}
        o_k(t)=\Big\{{\bf u}_k[n],{\bf q}_m[n],\Omega_k[n], Y_{k,m}^{\text{edge}}[n-1],\forall m \in \mathcal{M}\Big\}.
    \end{equation}
    \item \textbf{Action:} The decisions of MUs involve offloading association $\textbf{A}$, and offloading proportion $\bm{\varrho}$. Therefore, for each MU $k$, the action is decomposed by
    \begin{equation}
        a_k(t)=\Big\{\alpha_{k,m}[n],\rho_k[n],\forall m \in \mathcal{M}^*\cup\{0\}\Big\}.
    \end{equation}
    \item \textbf{Reward:} It can be noticed from $\eqref{obj}$ that the MUs need to consider their own influence on the total weighted energy consumption and the load on UAV servers. Hence, the reward of each MU $k$ should involve the energy consumption of both MU $k$ itself and the UAV associated by MU $k$. The reward of each MU agent $k$ is given by
    \begin{align}
    \nonumber
      r_k(t)=& -\sum\limits_{m=1}^M \alpha_{k,m}[n] \left(\varpi E_m^{\text{UAV}}[n]+P^{t}(T_k^{\text{loc}}[n],T_m^{\text{edge}}[n])\right) \\
     &-E_k^{\text{MU}}[n],
    \end{align}
    where 
    \begin{align}
    \nonumber
        P^t(T_k^{\text{loc}}[n],T_m^{\text{edge}}[n])=&\frac{\mu_t}{\delta_t}\Big(\text{ReLU}(T_k^{\text{loc}}[n]-\delta_t) \\
        &+ \text{ReLU}(T_m^{\text{edge}}[n]-\delta_t)\Big)
    \end{align}
    denotes the penalty for unsatisfaction of latency constraints, $\mu_t$ is a penalty factor, and the $\text{ReLU}(\cdot)$ is the rectified linear unit function.

\end{itemize}

\subsubsection{MDP element of UAVs}
Note that each UAV needs to make decision after MUs give their association. Herein, MDP elements are denoted as follows
\begin{itemize}
    \item \textbf{Observation:} The UAV agent $m$ observes the locations of themselves, locations of all MUs as well as the UAVs, and task information from associated MUs. Denoting $-m$ as the index set of UAVs except for UAV $m$, we have
    \begin{equation}
         o_{K+m}(t)=\Big\{\rho_k[n],\Omega_k[n], {\bf u}_k[n],{\bf q}_m[n],{\bf q}_{-m}[n]\Big\}.
    \end{equation}  
    \item \textbf{Action:} After receiving the requests, UAVs need to allocate the bandwidth, configure the computational frequency, and adjust velocity according to the observations. Therefore, the action of UAV agent $m$ is given by
    \begin{equation}
        a_{K+m}(t)=\bigg\{B_{k,m}[n],\tilde{f}_{k,m}^{\text{edge}}[n],{\bf a}_m[n],\forall k \in \mathcal{K}\bigg\}.
    \end{equation}
    \item \textbf{Reward:} After receiving, relaying and executing the tasks from MUs, the UAVs get rewards from environments. The reward of each UAV $m$ needs to consider the energy consumption of both itself and the served MUs, i.e.,
    \begin{align}
         r_{K+m}(t)=&\nonumber-\sum\limits_{k=1}^K \alpha_{k,m}[n]\left(E_k^{\text{MU}}[n]+P^t({T}_k^{\text{loc}}[n],{T}_m^{\text{edge}}[n])\right)  \nonumber\\
         &+\varpi E_m^{\text{UAV}}[n] +P^o({\bf q}_m[n])+P^c({\bf q}_m[n])  \nonumber\\
         &+P^d({\bf q}_m[n]),
    \end{align}
    where 
    \begin{align}
         P^o({\bf q}_m[n])=\mu_o\Vert{\bf q}_m[n]
        -\text{clip}({\bf q}_m[n],0,W)\Vert,
    \end{align}
    denotes the penalty when UAVs try to fly out of the square boundary with width $W$, and $\mu_o$ is a penalty factor.
    \begin{align}
        P^d({\bf q}_m[n])=&\frac{1}{W}\bigg(\Vert{\bf q}_m[n]-\frac{1}{\vert \mathcal{K}_m\vert}\sum\limits_{k=1}^K \alpha_{k,m}[n]{\bf w}_k[n]\Vert \nonumber\\
        & -d_{\rm th}\bigg)
    \end{align}
    guides the distance from UAV $m$ to MUs, $d_{\rm th}$ is a distance threshold, and $\vert \mathcal{K}_m\vert$ is the set of MUs associated with UAV $m$. In addition,
    \begin{align}
        & P^c({\bf q}_m[n])  \nonumber\\
        &=\mu_c\sum \limits_{j=1,j\neq m}^M \min\Big\{\Vert {\bf q}_m[n]-{\bf q}_j[n] \Vert - d_{\min},0\Big\}/d_{\min}
    \end{align}
    is the penalty for disobeying the safety distance $d_{\min}$ between UAVs, and $\mu_c$ is corresponding penalty
    factor. It is assumed that the UAVs will stop at the boundary if they try to fly out of it, and thus ${\bf q}_m[n] \leftarrow \text{clip}({\bf q}_m[n-1]+{\bf v}_m[n-1]\delta_t+\frac{1}{2}{\bf a}_m[n-1]\delta_t^2,0,W)$.
\end{itemize}

\begin{figure*}[t]
	\centering
	\includegraphics[width=0.95\textwidth]{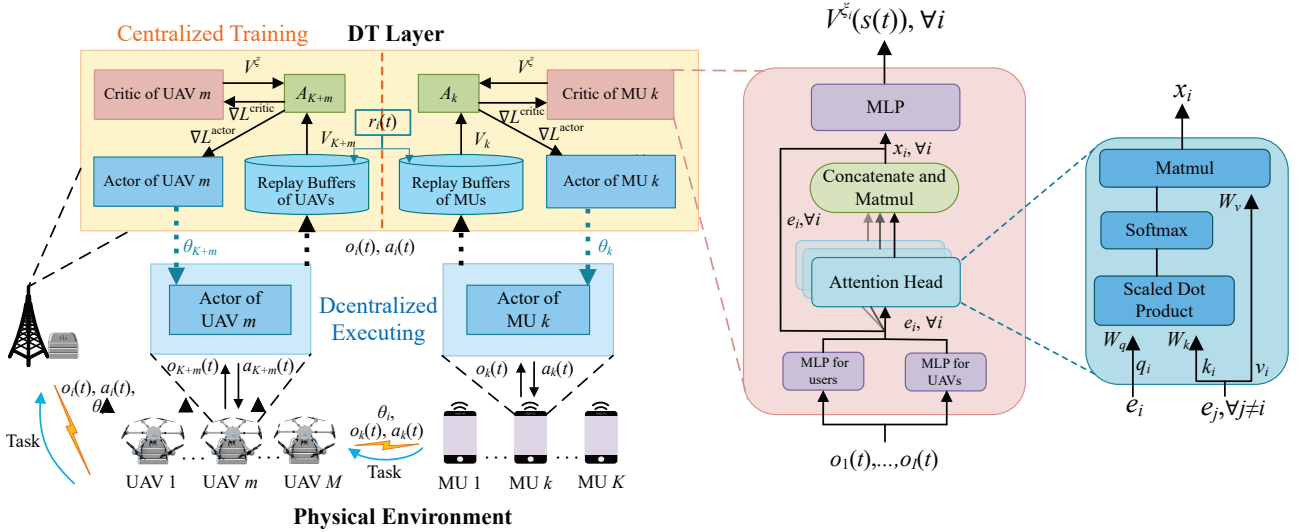}
	\caption{Training framework of AB-MAPPO.}
	\label{fig:mappo}    
\end{figure*}

\subsection{MAPPO-based DRL Approach With CTDE}
As a variant of PPO specialized for multi-agent settings, MAPPO is one of the state-of-the-art MARL algorithms \cite{Yu2021ar}. 
Based on on-policy scheme, each agent has an actor, a critic, and a replay buffer. 
As DT is capable of evaluating the global state of the environment, the centralized training and decentralized executing (CTDE) scheme can be adopted. 
In MAPPO with CTDE, the critics evaluate the centralized state-value function of global environment state. Specifically, the agents repeatedly input the observations $o_i(t)$ into actor networks and get $a_i(t)$ and $r_i(t)$ for an episode, storing the experiences into buffer. 
Then in the end of an episode, the agents update their policies. Firstly, they sample some batches of experiences $\{o_i(t),a_i(t),r_i(t),s(t),\text{pr}_i(t)\}, \forall i \in \mathcal{I}$ from their buffers, where the $\text{pr}_i(t)$ denotes the log-probability for sampling the action $a_i(t)$. Then, in each update, the actor and critic update the parameters with the loss of policy and global state-value, respectively. The loss of actor $i$ is calculated by

\begin{align}
     L^{\text{actor}}(\theta_i)=\mathbb{E}_{\pi_{\theta_i}}\Bigg\{\min\bigg[\frac{\pi_{\theta_i}(a_i(t)|o_i(t))}{\pi_{\theta_i^{'}}(a_i(t)|o_i(t))} \hat{A}_i(t), \nonumber\\
    \text{clip}\left(\frac{\pi_{\theta_i}(a_i(t)|o_i(t))}{\pi_{\theta_i^{'}}(a_i(t)|o_i(t))}, 1-\epsilon,1+\epsilon\right)\hat{A}_i(t)
    \bigg] \Bigg\},\label{eq:L_actor}
\end{align}
where $\pi_{\theta_i^{'}}$ and $\pi_{\theta_i}$ denote the old and current policy, respectively. $\hat{A}_i(t)$ is the estimation of advantage function $A_i(t)=Q_i(s(t),a_i(t))-V_i(s(t))$. In PPO series, the generalized advantage estimation (GAE) is adopted to improve the performance, which is defined as follows 
\begin{align}
    \hat{A}_i(t)=&\sum\limits_{l=0}^\infty(\gamma\lambda)^l\Big(R_i(t+l)+\gamma V_i\big(s(t+1+l)\big) \nonumber\\
    &-V_i(s(t+l))\Big),
\end{align}
where $\gamma$ is the discount factor, $\lambda$ is the parameter of GAE for bias-variance tradeoff in estimation, and $V_i(s(t))=\sum\limits_{l=0}^\infty\gamma^lR_i(t+l)$ is the cumulative discounted reward, which also represents the state-value function.
Denoting $V^{\xi_i}(s(t))$ as the state-value function estimated by the critic of agent $i$, the loss of critic $i$ is given by
\begin{equation}
    L_i^{\text{critic}}(\xi_i)=\frac{1}{2}\Big[V^{\xi_i}\big(s(t)\big)-V_i\big(s(t)\big)\Big]^2,\label{eq:L_critic}
\end{equation}
where $\xi_i$ is the parameter of $i$-th critic network. Therefore, the actor and critic can be updated according to \eqref{eq:L_actor} and \eqref{eq:L_critic}, respectively.

{It is worth noting that the constraints need to be satisfied by action-remapping as follows. First, the output vectors of actor networks in MAPPO are sampled from distributions and are divided according to the  sequence for concatenating the variables of actions. Then, the divided vectors are scaled into their origin domain. Furthermore, some constraints need to be tackled by normalization or rounding operation.  For each MU $k$, the index of its associated UAV is with respect to the maximum value of the divided vector corresponding to $\{\alpha_{k,m}[n],\forall m\in\mathcal{M}^*\cup\{0\}\}$, and thus $\{\alpha_{k,m}[n],\forall m\in \mathcal{M}\cup\{0\}\}$ is obtained. If $\alpha_{k,0}[n]=0$, $\alpha_{k,M+1}[n]$ will be decided by rounding its entry in the divided vector.   For constraints \eqref{c:band_alloc_on} and \eqref{c:rho_alpha} on  $B_{k,m}[n]$ and $\rho_k[n]$, we multiply their divided vectors by $[\alpha_{k,1}[n],\alpha_{k,2}[n],\ldots,\alpha_{k,M}[n]]$ and  $1-\alpha_{k,0}[n]$ as action masks. And then the constraints \eqref{c:band_alloc} and \eqref{c:f_max} are satisfied by normalizing their divided vectors.}

\subsection{MAPPO-based Training Framework}
 As displayed in Fig. \ref{fig:mappo}, during the proposed DT-assisted CTDE process, the MUs and UAVs perform computation offloading according to the actions given by the actor networks of their agents in physical environment, send their own experiences and synchronizing system status to the DT layer. 
 Then, the DT layer evaluates the global environment state by the observations of agents, updates the buffers, and gets the predicted values. 
 After updating the actors and critics, the parameters of actor networks are downloaded to UAVs and MUs. Note that the network parameters can be shared between homogeneous agents \cite{Yu2021ar}. 
 To fully exploit the performance of MAPPO approach, we introduce Beta distribution and attention, which are stated as below.
\subsubsection{Beta policy} The above-mentioned actions are typically continuous and bounded, such as ${\bf\varrho}\in[0,1]$ and ${\bf a}_k[n]\in[-a_{\max},a_{\max}]$. However, conventional action sampling from Gaussian distribution in policy networks will unavoidably introduce an estimation bias of policy gradient, since the boundary effects will be imposed by force clipping the values of out-of-bound actions. To tackle this problem, we adopt Beta distribution instead of Gaussian distribution in terms of the output of policy networks, which has the following form 
\begin{equation}
    f(x,\alpha,\beta)=\frac{\Gamma(\alpha+\beta)}{\Gamma(\alpha)\Gamma(\beta)}x^{\alpha-1}\Big(1-x\Big)^{\beta-1}, \label{eq:Beta}
\end{equation}
where $\alpha$ and $\beta$ are the parameters of Beta distribution. Since \eqref{eq:Beta} has also a bounded domain, it is appropriate to sample bounded actions. Moreover, the Beta distribution can be more close to uniform distribution than Gaussian by initializing, and thus the agents can more abundantly explore the action space at initial stage of training.

\begin{algorithm}[t]
	\caption{Proposed AB-MAPPO algorithm}
	\label{alg:mappo}
	\begin{algorithmic}[1]
		\STATE{Initialize $n=1$, episode length $\text{epl}$, PPO epochs $\text{Pe}$, and maximum episodes ${\rm e}^{\max}$.}
		\FOR {agent $i\in\mathcal{I}$}
		\STATE{Initialize actor networks $\theta_i$, critic networks $\xi_i$,replay buffer $\bm{B}_i$;}
		\ENDFOR
		\FOR {Episode = 1,$\dots$, ${\rm e}^{\max}$}
		\FOR {$t$ = 1,$\dots$, epl}
		\STATE {Obtain $o_k(t),\forall k \in \mathcal{K}$ from the environment;}
		\STATE {Execute the action $a_k(t)$, $\forall k \in \mathcal{K}$;}
		\STATE {Obtain $o_{K+m}(t),\forall m \in \mathcal{M}$ from the environment;}
		\STATE {Execute the action $a_{K+m}(t),\forall m \in \mathcal{M}$;}
		\STATE {Calculate log-probability $\text{pr}_{i}(t),\forall i \in \mathcal{I}$;}
        \STATE Synchronize status information to the DT layer if is required;
		\IF {$n$ mod $N$ = 0}
		\STATE Synchronize observations $o_i(t)$, actions $a_i(t)$ to the DT layer;
		\STATE {Evaluate reward $r_i(t),\forall i \in\mathcal{I}$ in DT layer;}
		\STATE {Store transition \\ $\text{Tr}_i(t)=$\{$o_i(t),a_i(t),r_i(t),s(t),\text{pr}_i(t)$\}, $\forall i \in \mathcal{I}$ into buffer $B_i$}; 
		\ENDIF
		\STATE{Update $n = n$ mod $N$ + 1}
		\ENDFOR
		\FOR {epoch = $1,\dots {\rm Pe}$}
		\FOR {agents $i\in\mathcal{I}$}
		\STATE {Update actor $\theta_i$ and critic $\xi_i$ according to \eqref{eq:L_actor} and \eqref{eq:L_critic} by $\forall \text{Tr}_i(t) \in \bm{B}_i$;}
		\ENDFOR
		\ENDFOR
		\STATE{Download the actor networks from DT layer to UAVs and MUs;}
		\ENDFOR
	\end{algorithmic}
\end{algorithm}

\subsubsection{Attention mechanism} With the increasing number of agents, the ever-increasing dimensions of environment state make it difficult for the critic network with a simple fully-connected layer to tackle the input, thereby leading to slower or even difficult convergence of critic network and being harmful to the behavior of actor network. {
However, for each agent, other agents may impose the impact with different intensities on the state-value, and thus the distinguished attention should be paid to them, which involves the nearby devices and the devices with higher computational load. 
Therefore, we introduce the multi-head attention unit before the multi-layer perceptron (MLP) of critic networks to 
improve the training process. The vectors of observations for two types of agents are first passed through their corresponding 3-layer MLPs to get feature values $e_i$. 
Then, all the feature values of agents are sent to the attention heads to get the attention values $x_i$ by}
\begin{equation}
    \alpha_{i,j}=\text{Softmax}\left(\frac{e_j^\text{T}  W_{\text{key}}^\text{T} W_q e_i }{\sqrt{d_{\text{key}}}}\right),
\end{equation}
\begin{equation}
    x_i=\sum\limits_{j\neq i}\alpha_{i,j} W_v e_j,
\end{equation}
where $e_j$ is the feature value of another agent $j$, $d_{\text{key}}$ is the variance of $e_j^\text{T}  W_{\text{key}}^\text{T} W_q e_i$. The matrix $W_{\text{key}}$ transforms $e_j$ into a key, the matrix $W_{v}$ transforms $e_j$ into a value, and the matrix $W_q$ transforms $e_i$ into a query.
Finally, $x_i$ and $o_i(t)$ are sent to the MLP to get the estimated state-value $V^{\xi_i}(s(t))$.
{
 \subsection{Complexity Analysis}
Based on above-mentioned discussions, we propose the AB-MAPPO algorithm, which is summarized in Algorithm \ref{alg:mappo}. The complexity of attention module is $\mathcal{O}(I^2V)$, where $V$ is the length of feature-value vectors, according to \cite{Gao2022TMC_Large}. For an MLP, the computational complexity of the $j$-th layer is $\mathcal{O}(Z_{j-1}Z_{j}+Z_{j}Z_{j+1})$, where $Z_j$ is the number of neurons for $j$-th layer. Hence, the computational complexity of a $J$-layer MLP is calculated by $\mathcal{O}\left( \sum_{j=2}^{J-1} Z_{j-1}Z_{j}+Z_{j}Z_{j+1} \right)$. The actor networks are all 3-layer MLPs, and each of two critic networks for MU and UAV agents have two 3-layer MLPs for the observations of different types agents and one 3-layer MLP after attention module for the output of value function. 
Therefore,  the overall computational complexity for training is calculated by the sum of complexity imposed by actor and critic networks $\mathcal{O}\left({\rm e}^{\max}({\rm Pe}I^2V + {\rm epl} \sum_{j=2}^{J-1} Z_{j-1}Z_{j}+Z_{j}Z_{j+1}) \right)$, and for one-step execution is just calculated by $\mathcal{O}\left(  \sum_{j=2}^{J-1} Z_{j-1}Z_{j}+Z_{j}Z_{j+1} \right)$.
}

\begin{table}[]
	\caption{{Environment Parameters}}
	\renewcommand{\arraystretch}{1.2}
	\begin{tabular}{|p{2.0in}|p{0.8in}|}
		\hline
		\textbf{Parameters}                       & \textbf{Value}  \\ \hline
		Channel bandwidth $B$                     & 50 MHz        \\ \hline       
		Channel power gain $\beta$                & -30 dB        \\ \hline       
		Environment parameters $a$ and $b$        & 15,0.5         \\ \hline      
		Time period $T$                           & 60 s          \\ \hline       
		Time slot $\delta_t$                      & 1 s           \\ \hline      
		Transmit power of MUs $p_k^{\text{MU}}$   & 0.2 W         \\ \hline      
		Transmit power of UAVs $p_m^{\text{UAV}}$ & 0.5 W         \\ \hline       
		Path loss exponent $\tilde{\iota}$        & 2.2         \\ \hline         
		NLoS attenuation $\nu$                    & 0.2          \\ \hline       
		Noise power density $N_0$                                 & -127 dBm/Hz         \\ \hline
		Effective capacitance coefficient $\kappa$                & $10^{-27}$                           \\ \hline
		Maximum CPU frequency of MU $k$ $f_{\max}^{\text{loc}}$   & 1 GHz                                \\ \hline
		Maximum CPU frequency of UAV $m$ $f_{\max}^{\text{edge}}$ & 10 GHz                               \\ \hline
		Maximum velocity of UAV $v_{\max}$                        & 30 m/s                               \\ \hline
		Maximum acceleration $a_{\max}$                           & 5 $\text{m/s}^2$                     \\ \hline
		UAV settings $P_0,P_i$                                    & 39.04 W, 79.07 W                     \\ \hline
		UAV settings $U_{\text{tip}},v_0,A$                       & 120 m/s, 3.6 m/s, 0.5030 $\text{m}^2$ \\ \hline
		Weight factor $\varpi$                                    & 0.001                                        
		\\ \hline
	\end{tabular}
	\label{tab:sim_param}
\end{table}

\section{Numerical Results}\label{s:simulation}

In this section, we evaluate the performance of the proposed AB-MAPPO algorithm. The simulation settings and numerical results are presented as follows.

\subsection{Simulation Settings}
In our simulations, a 1000 $\times$ 1000 m rectangular area is considered, where the MUs are randomly distributed in region with $x,y\in[0,1000]$ m, the BS is set at (-500 m, 0 m, 10 m), and the UAVs are flying at $H=200$ m. Note that since the energy consumption of UAVs and MUs are sensitive and the BS typically has abundant power supply, the energy consumption of the BS involving computation and DT construction is not considered.
{Other parameter settings of simulation and hyperparameters of MAPPO are summarized in Table \ref{tab:sim_param} and Table \ref{tab:sim_param_MAPPO}, according to prior works \cite{Wang2022TMC,Dai2022TVT,Hu2020TCOMM,Zeng2019TWC}.}

The baseline algorithms for comparison with our proposed AB-MAPPO are listed as follows:
\begin{itemize}
    \item \textbf{B-MAPPO:} MAPPO without attention mechanism and with Beta distribution.
    \item \textbf{AG-MAPPO:} MAPPO with Gaussian distribution and attention mechanism.
    \item \textbf{MADDPG:} MADDPG is an off-policy algorithm with determinate policy \cite{Wang2021TCCN}. The exploration noise is set as 0.5, the buffer size is set as 20000, and other parameters are the same to the proposed scheme.
    \item \textbf{Randomized:} The MUs and UAVs execute randomly generated actions by the algorithm. 
\end{itemize}

\begin{table}
	\caption{{Hyperparameters Parameters of AB-MAPPO}}\centering
	 \renewcommand{\arraystretch}{1.2}
	\begin{tabular}{|l|l|}
		\hline
		\textbf{Parameters}                           & \textbf{Value}   \\ \hline
		Total number of steps                         & 80k             \\ \hline
		Length of an episode $\text{epl}$             & 300             \\ \hline
		Learning rate of actor                        & 0.0003          \\ \hline
		{Discount factor of UAVs}                     & 0.95            \\ \hline
		Discount factor of MUs                        & 0.8            \\ \hline
		Penalty factors $\mu_o$, $\mu_t$, and $\mu_c$ & 0.1            \\ \hline 
		Number of heads in attention unit             & 4              \\ \hline
		Number of mini-batches                            & 1              \\ \hline
		PPO epoch $\text{Pe}$                             & 5              \\ \hline
		Learning rate of critic                           & 0.0003         \\ \hline
		{Number of hidden layers of MLPs}                 & 1              \\ \hline
		{Hidden size of MLPs}                            & 128            \\ \hline
		Optimizer                                         & Adam           \\ \hline
		$d_{\rm th}$                                     & 300 m          \\ \hline
	\end{tabular}
	\label{tab:sim_param_MAPPO}
\end{table}

\subsection{Convergence of the MARL Training Algorithm}
\begin{figure}[t]
	\centering  
	\subfigure[The average reward of MUs.]{
        \begin{minipage}{0.45\textwidth}
            \includegraphics[width=3.2in]{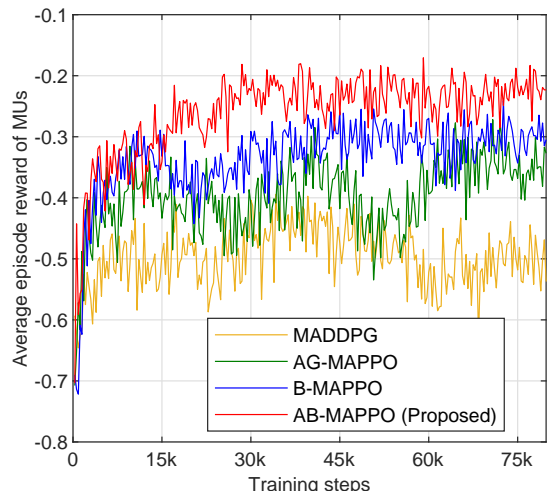}
        \end{minipage}
    }
	\subfigure[The average reward of UAVs.]{
        \begin{minipage}{0.45\textwidth}
            \includegraphics[width=3.2in]{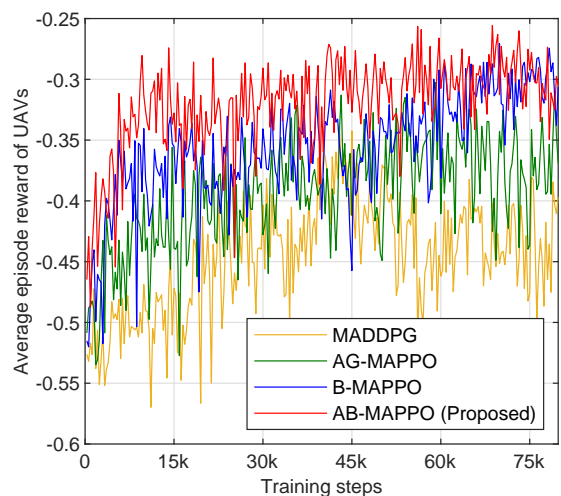}
        \end{minipage}
    }
    \caption{The convergence of the rewards of agents.}
    \label{fig:convergence}
\end{figure}
{
We first evaluate the convergence performance of the proposed AB-MAPPO algorithm in Fig. \ref{fig:convergence}(a) and Fig. \ref{fig:convergence}(b). The average episode reward of all MUs is evaluated with comparison of RL baselines under $M=10$ UAVs and $K=60$ MUs. It can be seen that the average episode reward of UAVs obviously increases during the training process and then reaches to the stable values. Intuitively, the proposed AB-MAPPO scheme, B-MAPPO scheme, AG-MAPPO scheme, and MADDPG scheme converge at around 30k, 45k, 60k and 60k steps, respectively. As expected, the proposed scheme outperforms other baselines, with the fastest convergence and reward. Specifically, the proposed AB-MAPPO scheme converge faster than the B-MAPPPO scheme, and the MAPPO schemes with Beta distribution achieve greater rewards than the AG-MAPPO scheme. The reason can be explained as follows: 1) The attention mechanism makes the critic networks quickly concentrate on the significant parts of the global state, thereby accelerating the convergence. 2) With the Beta distribution, the agents explore more uniformly at the initial stage of training, retain better exploration ability, and converge to better solutions, which also presents the superiority of Beta distribution over Gaussian distribution and the MADDPG's exploration noise. 
}

\begin{figure}[t]
	\centering
	
	\begin{minipage}[t]{0.45\textwidth}
        \centerline{\includegraphics[width=3.2in]{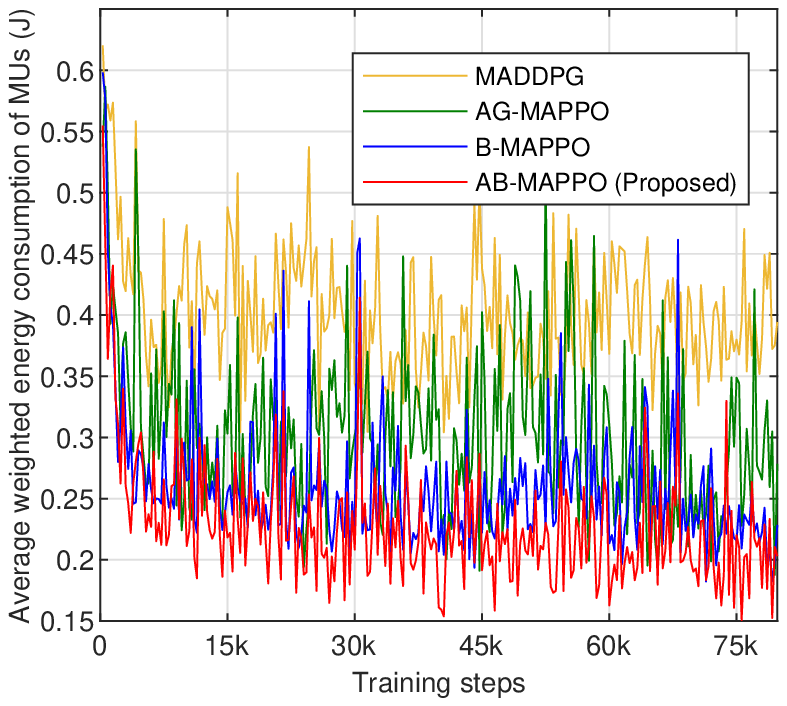}}
        \caption{The convergence of weighted energy consumption.}
        \label{fig:weightedEnergy}
	\end{minipage}
	\begin{minipage}[t]{0.45\textwidth}
        \centerline{\includegraphics[width=3.2in]{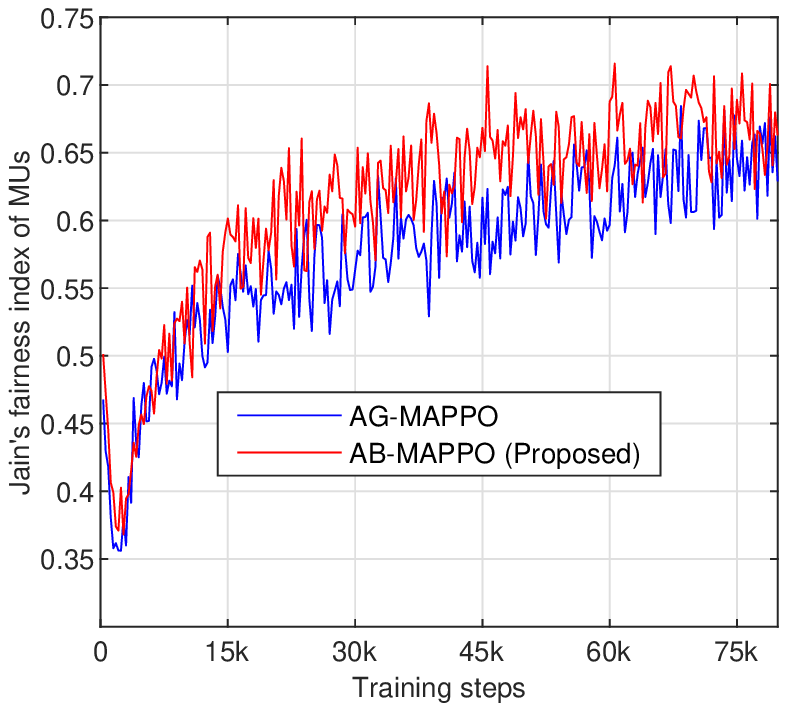}}
        \caption{The improvement of fairness of MUs.}
        \label{fig:Jain}
	\end{minipage}
	
\end{figure}

{
Subsequently, Fig. \ref{fig:weightedEnergy} presents the convergence of the average weighted energy consumption of users with $K=60$ MUs and $M=10$ UAVs. It indicates that the energy consumption has also been optimized with the increase of reward during the training process. Intuitively, the proposed AB-MAPPO scheme achieves lower energy consumption and has the most rapid convergence. In terms of stability, the energy consumption of the proposed AB-MAPPO scheme goes down more smoothly, the AG-MAPPO and the B-MAPPO schemes have higher fluctuation during training, and the MADDPG scheme is more tortuous. The above results verify the effectiveness and reliability of our proposed scheme. 
}

{
To verify the fairness of service, we present the varying of Jain's fairness index of MUs in Fig. \ref{fig:Jain} during training. It is worth noting that the Jain's fairness index of service is denoted as $\frac{\sum\limits_{k=1}^K E_k^{\text{MU}}[n])^2}{(K\sum\limits_{k=1}^K (E_k^{\text{MU}}[n])^2} $. We can find that the fairness of MUs gradually grows during the training process and the fairness of proposed AB-MAPPO is higher and more stable against the AG-MAPPO scheme. Another observation is that the fairness of both schemes slightly decreases at first and then ascends, while the AG-MAPPO scheme converges faster but has lower fairness. The ascending of fairness is due to the fact that the policy of MUs gradually becomes stable, indicating that both the quality and reliability of service have been improved. 
}

\subsection{Comparison of Benchmarks Under Different Settings}

\begin{figure}[t]
	\centering
	
	\begin{minipage}[t]{0.45\textwidth}
        \centerline{\includegraphics[width=3.2in]{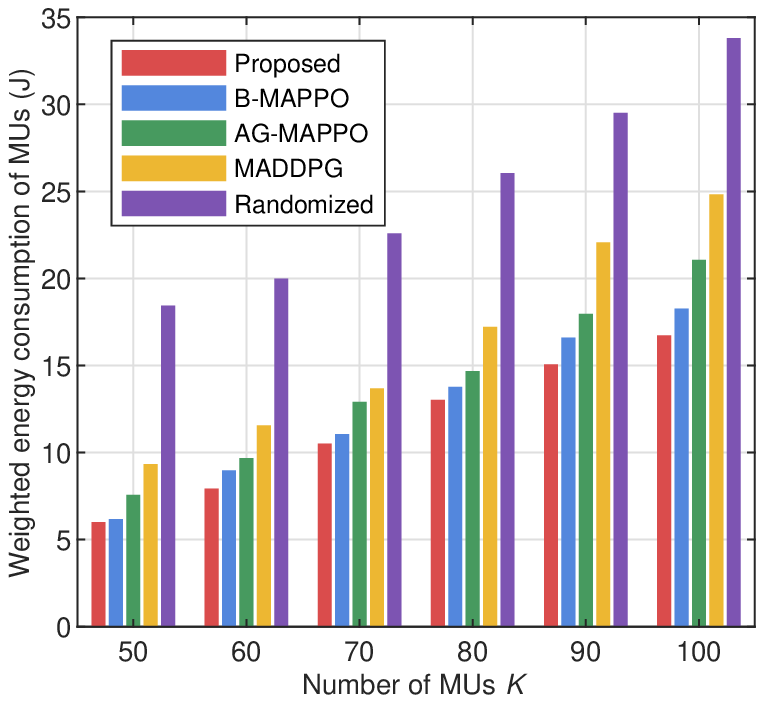}}
        \caption{The impact of MUs on average weighted energy consumption of MUs.}
        \label{fig:numberofMUs}
	\end{minipage}
	\begin{minipage}[t]{0.45\textwidth}
        \centerline{\includegraphics[width=3.2in]{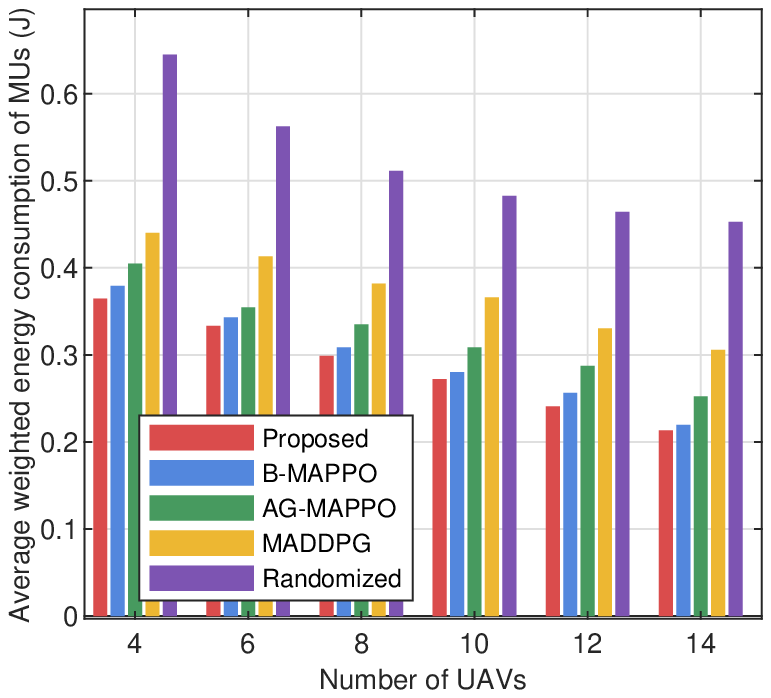}}
        \caption{The impact of number of UAVs on average weighted energy consumption of MUs.}
        \label{fig:numberofUAVs}
	\end{minipage}
	
\end{figure}

{
To verify the impact of number of MUs on the energy expense of the network, we then evaluate the average weighted energy consumption of MUs, i.e., $\sum\limits_{k=1}^K\left(\varpi \sum\limits_{m=1}^M \alpha_{k,m}[n] E_m^{\text{UAV}}[n]+E_k^{\text{loc}}[n]\right)/K$, versus the number of MUs in Fig. \ref{fig:numberofMUs} with $M=10$ UAVs. From this figure,  we observe that the energy cost gradually increases as the number of MUs grows, the proposed AB-MAPPO scheme outperforms the baselines. Additionally, the gaps between RL-based schemes have the tendency of becoming larger, the randomized scheme keeps at the highest energy consumption, and the performance gain of attention mechanism increases as the number of MUs grows.  Since the average communication resource declines as the number of MUs grows, the latency and energy for transmission of MUs increase, and the MEC servers can process less amount of tasks. Therefore, the MUs tend to computing locally, thereby leading to the ascending of the energy for local computing. Furthermore, the randomized scheme uniformly generates actions, and thus has poor ability to appropriately utilize the abundant resource at BS and UAVs.
}

\begin{figure}[t]
	\centering
	
	\begin{minipage}[t]{0.45\textwidth}
        \centerline{\includegraphics[width=3.2in]{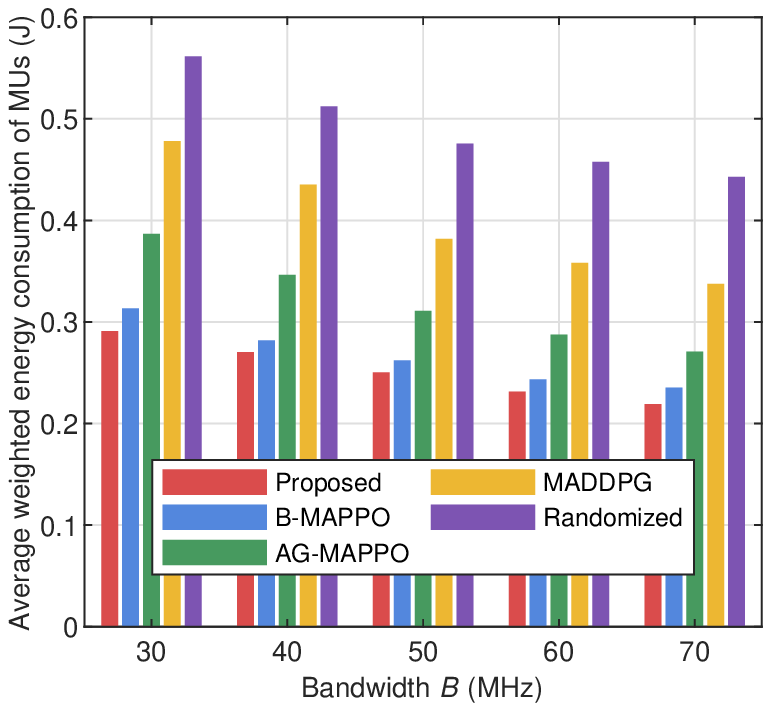}}
        \caption{The impact of bandwidth on energy consumption of MUs.}
        \label{fig:bandwidth}
	\end{minipage}
	\begin{minipage}[t]{0.45\textwidth}
        \centerline{\includegraphics[width=3.2in]{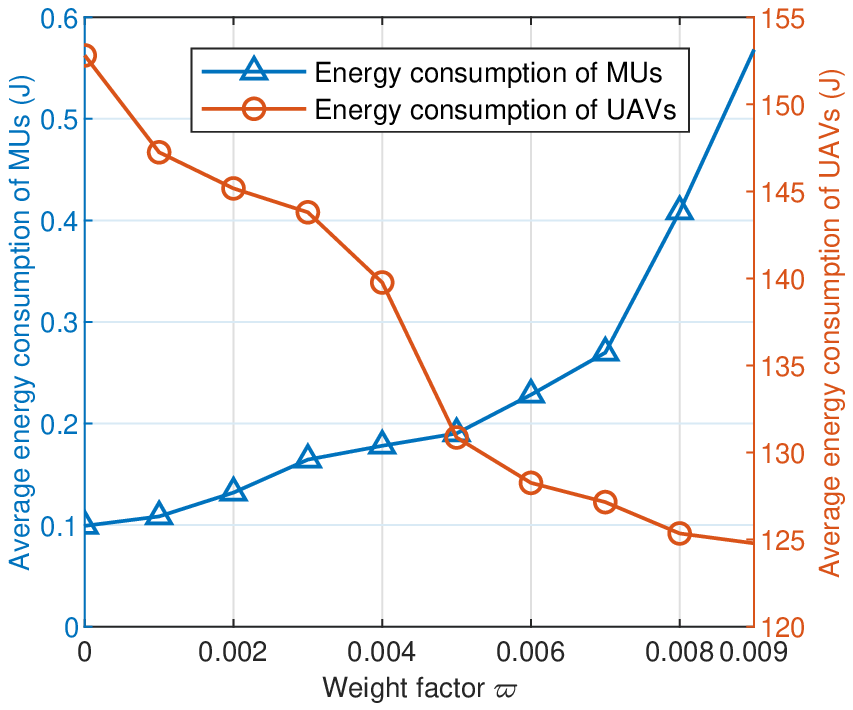}}
        \caption{The impact of weight factor on energy consumption of MUs and UAVs.}
        \label{fig:weight}
	\end{minipage}
	
\end{figure}

{
Fig. \ref{fig:numberofUAVs} presents the comparison of the proposed scheme and baselines on average weighted energy consumption of $K=70$ MUs under different number of UAVs. As shown in this figure, when more UAVs participate in the service, the average weighted energy of MUs gradually declines. It can be seen that the proposed AB-MAPPO scheme has the lowest energy cost, and the B-MAPPO scheme is slightly higher than it. The likely reason is that the Beta distribution can efficiently improve the exploration of UAVs in large action space. Moreover, the MADDPG scheme converges to worse solutions than that of the MAPPO scheme, which indicates that the algorithm is relatively difficult to tackle the increasingly challenging actions.
}

\begin{figure}[t]
	\centering  
	{
		\begin{minipage}[t]{0.22\textwidth}
			\centerline{\includegraphics[width=1.7in]{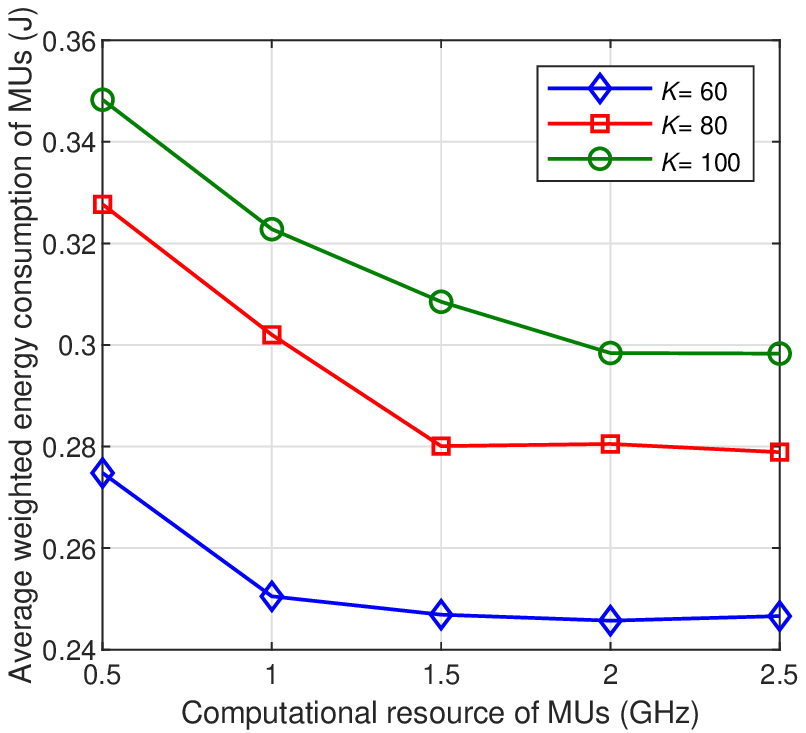}}
		\end{minipage}
	}
	{
		\begin{minipage}[t]{0.22\textwidth}
			\centerline{\includegraphics[width=1.7in]{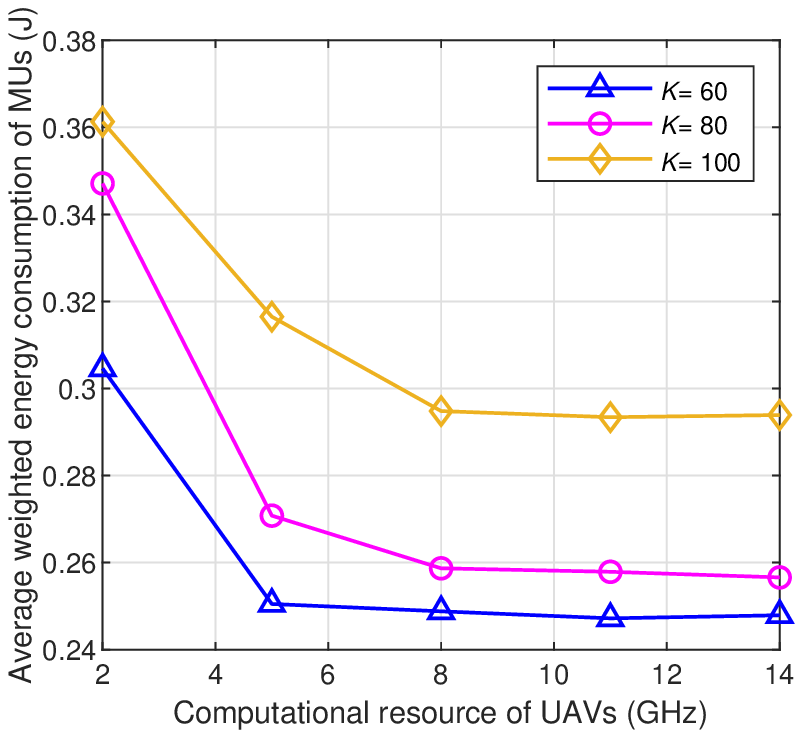}}
		\end{minipage}
	}
	\caption{The impact of computational resource.}
	\label{fig:f}
\end{figure}

{
To gain further insights, Fig. \ref{fig:bandwidth} depicts the average weighted energy consumption of MUs versus the channel bandwidth $B$. It can be seen that the energy consumption gradually decreases when the bandwidth increases. Furthermore, the proposed AB-MAPPO scheme obtains the best performance, and when the bandwidth becomes larger, the gaps between RL schemes have the tendency of decreasing. This is because more abundant communication resource enables the MUs to offload more tasks to UAVs and BS, since lager bandwidth imposes looser constraints on the optimization problem, which may introduce larger penalty to hinder the exploration of agents. This in turn verifies that the proposed AB-MAPPO scheme outperforms the baselines in tackling the problems with strong constraints.
}

\subsection{The Impact of Environment Settings on the Performance}

\begin{figure}[t]
		\centerline{\includegraphics[width=3.2in]{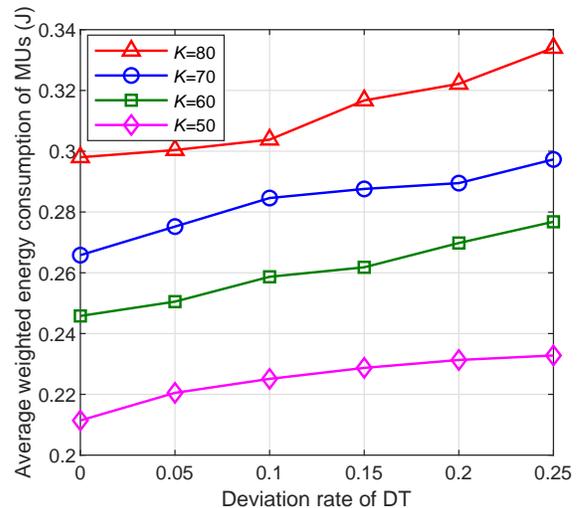}}
		\caption{The impact of DT deviation.}
		\label{fig:DT_bias}
\end{figure}

In Fig. \ref{fig:weight}, we examine the impact of weight factor $\varpi$ on the optimized energy consumption of UAVs and MUs, {under $K=60$ MUs and $M=10$ UAVs}. It can be observed that as the weight factor $\varpi$ increases from 0 to 0.009, the energy consumption of UAVs sharply decreases at first, and then becomes smoother. Meanwhile, the energy of MUs increases sharply. The reason for this trend is that as $\varpi$ increases, the energy consumption of UAVs is more emphasized and more tasks are computed locally by MUs, which means that more computing energy is consumed by MUs and less energy is consumed by UAVs.

To evaluate the impact of computational resource on the energy expense of users, we present the average weighted energy consumption of users versus the maximum computational frequency of users and UAVs in Fig. \ref{fig:f}(a) and Fig. \ref{fig:f}(b), respectively. It can be readily observed that  as the computational frequency of users and UAVs increases, the energy consumption gradually decreases, and keeps at about constant values. The likely reason is that the increase of computational resource leads to an increase of satisfaction with computational service. Hence, the policy of MUs on offloading proportion is accordingly adjusted to balance the energy consumption and the penalty for the improvement of reward. Then, the abundant computational resource makes the latency requirements satisfied, and the policies become stable.

\begin{figure}[t]
	\centerline{\includegraphics[width=3.2in]{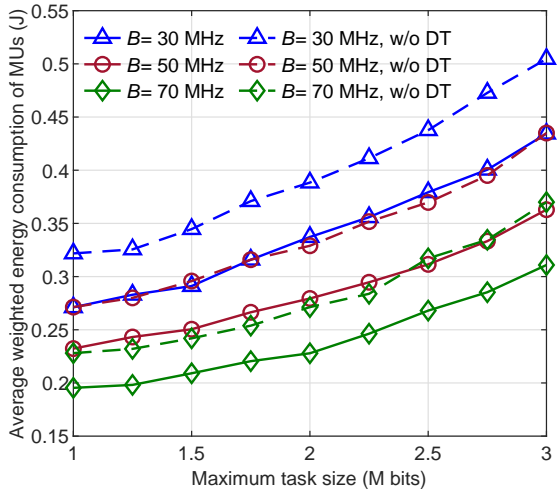}}
	\caption{The impact of maximum task size of users.}
	\label{fig:taskSize}
\end{figure}

We then evaluate the deviation of DT on system performance in Fig. \ref{fig:DT_bias}. We can find that as the deviation rate of DT increases, the average energy consumption of MUs has the tendency of increasing. This can be explained by the fact that the DT deviation reduces the accuracy of observations and the actions of agents. As such, the improvement of the policies for agents is influenced, and thus the reward and the energy consumption become worse. Another observation is that as the deviation of DT increases from 0 to 0.25, the weight energy consumption rises by about 15\%, which verifies the robustness of the proposed MARL approach.

\begin{figure}[t]
	\centering  
	\subfigure[Scenario 1 with $K=12$, $M=3$, and $T=50$ s]{
		\begin{minipage}[t]{0.21\textwidth}
			\includegraphics[width=1.5in]{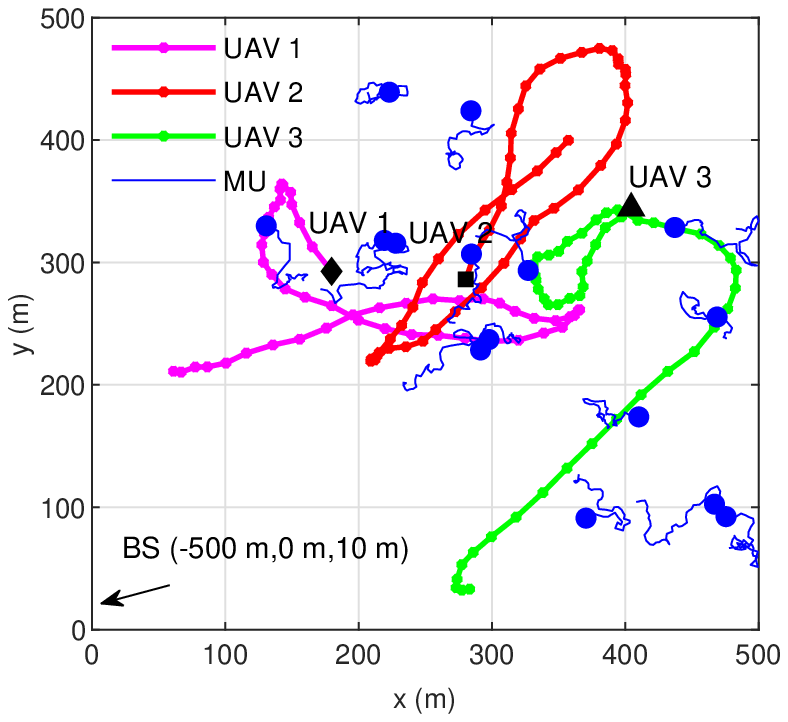}
		\end{minipage}
	}
	\subfigure[Scenario 2 with $K=12$, $M=3$, and $T=50$ s]{
		\begin{minipage}[t]{0.21\textwidth}
			\includegraphics[width=1.5in]{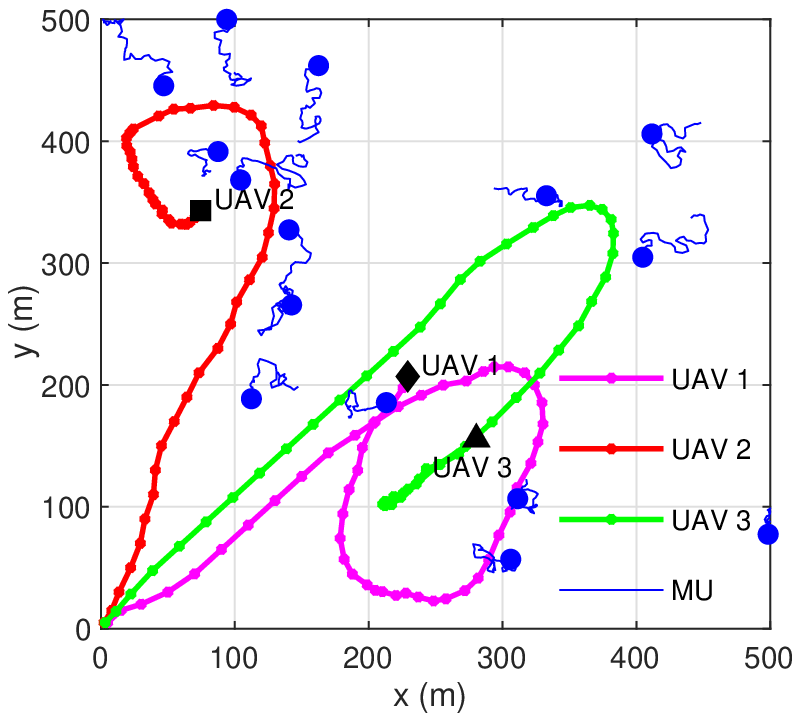}
		\end{minipage}
	}
	
	\subfigure[Scenario 3 with $K=60$, $M=10$, and $T=60$ s]{
		\begin{minipage}[t]{0.21\textwidth}
			\includegraphics[width=1.5in]{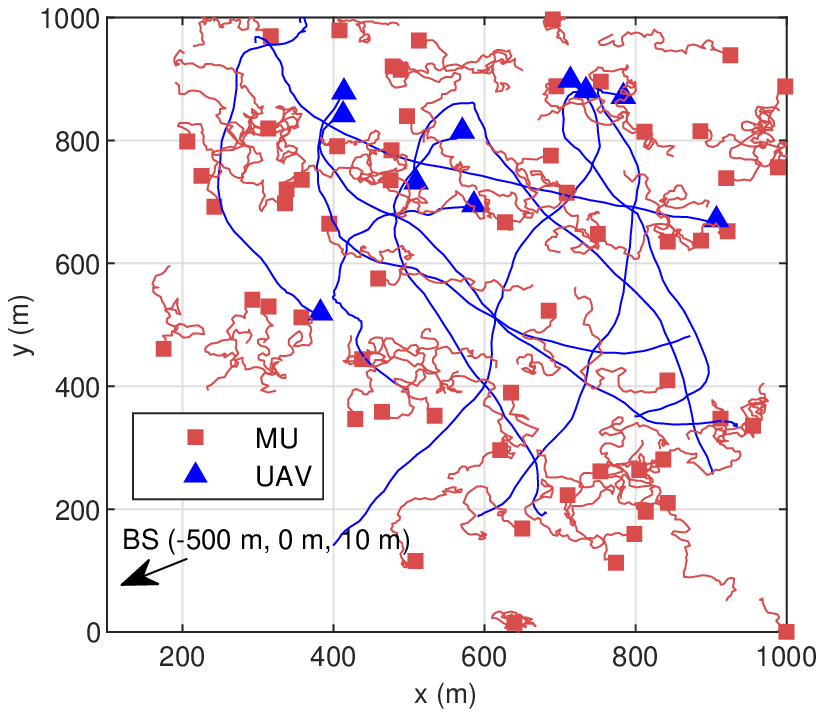}
		\end{minipage}
	}
	\subfigure[Scenario 4 with $K=60$, $M=10$, and $T=60$ s]{
		\begin{minipage}[t]{0.21\textwidth}
			\includegraphics[width=1.5in]{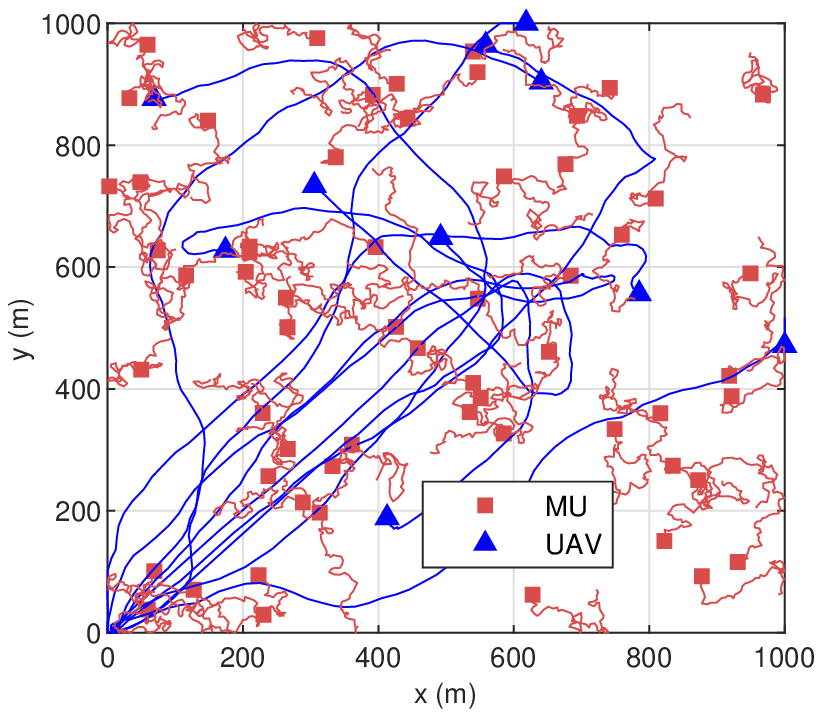}
		\end{minipage}
	}
	\caption{The trajectories of UAVs with different scenarios.}
	\label{fig:trajectory}
\end{figure}

Fig. \ref{fig:taskSize} shows weighted energy consumption versus the maximum task size $D_{\max}$ and evaluates the without DT scheme (``labeled as w/o DT''), while fixing the minimum value $5\times10^5$ for with different number of MUs. The  w/o DT scheme is evaluated by adding the random deviation rate in [-0.5,0.5] on the normalized variables of locations and computational frequency. This is under the following consideration. The DT-assisted framework enables UAVs to directly synchronize the estimated information with the DT layer at BS. Hence, the estimated deviation is cut down by avoiding the frequent queries for exchanging the environment information between UAVs, and thus the devices can obtain more timely status of the system. Therefore, the estimated deviation is set higher to evaluate the w/o DT schemes, whose performance is intuitively worse than that of DT-assisted schemes. Additionally, it can be readily observed that more energy is consumed as the task size grows and the bandwidth decreases. It can be explained by the fact that more computation resource is required to satisfy the latency requirements of MUs, and since the bandwidth decreases, the average communication resource decreases. As such, the transmission expense ascends and the MUs tend to compute more tasks locally, thereby leading to the increasing of energy consumption.

Fig. \ref{fig:trajectory} displays the trajectories of MUs and UAVs in different scenarios. In Fig. \ref{fig:trajectory} (a), the MUs are relatively crowded in a part of region, and the UAVs start from random locations, with the width of region 500 m and $T=50$ s. It can be observed that the trained UAVs are capable of flying closer to MUs and keep hovering on the crowded MUs to pursue higher transmit rate. 
In Fig. \ref{fig:trajectory} (b), the UAVs start at the corner of the region, and the MUs are relatively dispersed. At the beginning, the UAVs can quickly fly to the MUs, and then cooperate with each other to serve the associated MUs. Additionally, the UAVs are close to most of associated MUs rather than fully involving the remote ones, which indicates that UAVs can skillfully the long-term reward.
In Figs. \ref{fig:trajectory} (c) and (d), we present the trajectory with large-scale simulations with $K=60$ users and $M=10$ UAVs. It can be seen that the UAVs can move to the regions with more MUs and cooperatively cover the region of MUs to provide better service and maintain the fairness. It verifies that in the situation with larger scale, the reward function can guide the UAVs to rapidly access the MUs.

\section{Conclusion}\label{s:conclusion}
In this paper, we proposed a multi-UAV-assisted MEC network with air-ground cooperation where DT is applied to enhance the offloading service. We formulated a weighted sum energy consumption minimization
problem by jointly optimizing the offloading decision, bandwidth, flying trajectory, communication resource and computation resource. To tackle this challenging problem, we modeled our problem as an MDP where MUs and UAVs act as agents to collectively interact with the environment to receive distinctive observations. Considering the high-dimensional hybrid action space, the MAPPO algorithm with attention mechanism and Beta distribution was leveraged to efficiently obtain the optimal policy. Simulation results verified that the proposed scheme can significantly reduce energy consumption of the network compared with other benchmarks. Additionally, the combination of UAV and BS offloading strategies can  take fully advantage of communication and computational resources, as well as adapt to the time-varying network environments.

\bibliographystyle{IEEEtran}
\bibliography{IEEEabrv,refs}

\end{document}